\documentstyle[12pt,epsfig,cite]{article}

\newlength{\dinwidth}
\newlength{\dinmargin}
\def\docnum#1{\hbox to \hsize{\hskip123mm\hbox{#1}\hss}}
\def\date#1{\edef\@temp{#1}\ifx\@temp\@empty\def\@temp{\today}\fi
\hbox to \hsize{\hskip123mm\hbox{\@temp}\hss}}
\def\title#1{\vskip 0.8in plus 2in\begin{center}%
{\Large\bf#1\par}\vskip1.5em\end{center}\vskip 1in}
\def\@makefnmark{\hbox{$^{\@thefnmark)}$}}
\def\author#1{
\setcounter{footnote}{0}\def\@currentlabel{}%
\begingroup\def\thefootnote{\arabic{footnote}}
\def\@makefnmark{\hbox{$^{\@thefnmark)}$}}
\global\@topnum\z@ \large\begin{center}{\lineskip.5em
\begin{tabular}[t]{c}#1\end{tabular}\par}
\end{center}\par\vskip1.5em\@thanks\endgroup}

\def\abstract{\vskip0.8in plus 3in\begin{center}{\large\bf Abstract}\end{center}\quotation}

\newcommand{\QG}   {{\bf{Q}}}
\newcommand{\QGz}  {{\bf{Q}}^0}
\newcommand{\QGzi} {{\bf{Q}}_i^0}
\newcommand{\QGzj} {\vec{Q}_j^0}
\newcommand{\QGi}  {{\bf{Q}}_i}
\newcommand{\PG}   {{\bf \phi}}
\newcommand{\betb} {{\bf \beta}}
\newcommand{\ppb}  {$\rm{p\bar{p}}\;$}
\newcommand{\cc}   {\rm{c{\bar c}}}
\newcommand{\bb}   {\rm{b{\bar b}}}
\newcommand{\ee}   {e$^+$e$^-$}
\newcommand{\eeqq} {$\rm{e^+e^- \rightarrow q \bar{q}}\;$}
\newcommand{\qi}   {{\bf{q}}_i}
\newcommand{\qj}   {{\bf{q}}_j}
\newcommand{\ql}   {{\bf{q}}_l}
\newcommand{\qk}   {{\bf{q}}_k}

\newcommand{\zum}  {{\rm{\Sigma}}}
\newcommand{\dint} {{\rm{d}}}
\newcommand{\pa}   {{\rm{(particles)}}}
\newcommand{\E}    {{\rm{e}}}
\newcommand{\I}    {{\rm{i}}}

\newcommand{\ppbf} {${\bf{p\bar{p}}}\;$}

\begin{document}

\begin{titlepage}
\flushright{DFF 268/02/1997}
\flushright{February 1997}
\title{Thermal hadron production in pp and \ppbf collisions} 
\vspace{-2.0cm}
\centerline{\large{F. Becattini}} 
\vspace{0.5cm} 
\centerline{\it{INFN Sezione di Firenze}}  
\centerline{\it{Largo E. Fermi 2, I-50125 Firenze}} 
\centerline{e-mail: becattini@fi.infn.it}
\vspace*{1cm} 
\centerline{\large {and U. Heinz}\footnote{supported in part by BMBF, DFG and GSI}} 
\vspace{0.5cm}
\centerline{\it {Institut f\"ur Theoretische Physik, Universit\"at Regensburg}} 
\centerline{\it {D-93040 Regensburg, Germany}}
\centerline{e-mail: Ulrich.Heinz@physik.uni-regensburg.de}

\begin{abstract}
 It is shown that the hadron production in high energy pp 
 and \ppb collisions, 
 calculated by assuming that particles originate in hadron gas fireballs
 at thermal and partial chemical equilibrium, agrees very well with the data. 
 The temperature of the hadron gas fireballs, determined by fitting hadron 
 abundances, does not seem to depend on the centre of mass energy, having a 
 nearly constant value of about 170 MeV. This value is in agreement with that 
 obtained in \ee collisions and supports a universal hadronization mechanism 
 in all kinds of reactions consisting in a parton-hadron transition at critical
 values of temperature and pressure.  
\end{abstract}
\vspace*{1.5cm}
\centerline{\it{Published in Z. Phys. C}}

\end{titlepage}

\section{Introduction}
 
 The thermodynamic approach to hadron production in hadronic 
 collisions was originally introduced by Hagedorn \cite{hag} about 
 thirty years ago. The most important phenomenological indication of 
 thermal multihadron production in high energy reactions was found in 
 the universal slope of the transverse mass (i.e. $m_T=\sqrt{p^2_T+m^2}$) 
 spectra \cite{hag2}, where transverse means orthogonal to the beam 
 line. This kind of signature of a hadron gas formation is nowadays 
 extensively used in heavy ions reactions, although it has been 
 realized that transverse collective motion of the hadron gas may 
 significantly distort the basic thermal $m_T$-spectrum 
 \cite{heinz}, thus complicating the extraction of the temperature. A 
 much better probe of the existence of locally thermalized sources in 
 hadronic collisions is the overall production rate of individual 
 hadron species which, being a Lorentz-invariant quantity, is not 
 affected by local collective motions of the hadron gas. However, the 
 analysis of hadron production rates with the thermodynamical {\it 
 ansatz} implies that inter-species {\it chemical} equilibrium is 
 attained, which is a much tighter requirement than that of {\it 
 thermal-kinetic} intra-species equilibrium assumed in the analysis of 
 $m_T$ spectra \cite{heinz2}. Chemical equilibrium thus usually 
 implies also thermal kinetic equilibrium. For this reason we focus 
 our attention in this paper on the analysis of hadron abundances and 
 the question of chemical equilibrium, leaving the analysis of 
 $m_T$-spectra (with its possible complications due to collective 
 dynamical effects) to a separate publication. \\
 The smallness of the collision systems studied here requires 
 appropriate theoretical tools: in order to properly compare 
 theoretical predicted multiplicities to experimental ones, the use of 
 statistical mechanics in its canonical form is mandatory, that means 
 exact quantum numbers conservation is required, unlike in the
 grand-canonical formalism \cite{hag3}. It will be shown indeed that 
 particle average particle multiplicities in small systems are heavily 
 affected by conservation laws well beyond what the use of chemical 
 potentials predicts (this was previously observed in a similar 
 canonical thermodynamic analysis of \ppb annihilation at rest 
 \cite{bluemel}). However, in the high multiplicity (or large
 volume) limit the grand-canonical formalism recovers its validity.  
 This paper generalizes the thermodynamical model introduced in ref. 
 \cite{beca} for \ee collisions by releasing some assumptions which 
 were made there; calculations are performed with a larger symmetry 
 group (actually by also taking into account the conservation of the
 electric charge). Moreover, formulae of global correlations between 
 different particles species are provided, and a comparison with data 
 is made in this regard as well.

\section{The model}
 
 In refs. \cite{beca,tesi} a thermodynamical model of hadron production in 
 \ee collisions was developed on the basis of the following 
 assumption: the hadronic jets observed in the final state of a \eeqq 
 event must be identified with hadron gas phases having a collective 
 motion. This identification is valid at the decoupling time, when 
 hadrons stop interacting after their formation and (possibly) a short 
 expansion ({\em freeze-out}). Throughout this paper we will refer to 
 such hadron gas phases with a collective motion as {\em fireballs}, 
 following refs. \cite{hag,hag2}. Since most events in a \eeqq 
 reaction are two-jet events, it was assumed that two fireballs are 
 formed and that their internal properties, namely quantum numbers, 
 are related to those of the corresponding primary quarks. In the 
 so-called {\em correlated jet scheme} correlations between the 
 quantum numbers of the two fireballs were allowed beyond the simple 
 correspondance between the fireball and the parent quark quantum 
 numbers. This scheme turned out to be in better agreement with the 
 data than a correlation-free scheme \cite{beca}. \\ 
 The more complicated structure of a hadronic collision does not allow 
 a straightforward extension of this model. If the assumption of 
 hadron gas fireballs is maintained, the possibility of an arbitrary number 
 of fireballs with an arbitrary configuration of quantum numbers 
 should be taken into account \cite{faro}. To be specific, let us define
 a vector $\QG = (Q,N,S,C,B)$ with integer components equal to the electric 
 charge, baryon number, strangeness, charm and beauty respectively. 
 We assume that 
 the final state of a pp or a \ppb interaction consists of a set of 
 $N$ fireballs, each with its own four-vector $\beta_i=u_i/T_i$, where 
 $T_i$ is the temperature and $u_i= (\gamma_i,\betb_i \gamma_i)$ is 
 the four-velocity \cite{tous}, quantum numbers $\QGzi$ and volume in 
 the rest frame $V_i$. The quantum vectors $\QGzi$ must fulfill the 
 overall conservation constraint $\sum_{i=1}^N \QGzi = \QGz $ where 
 $\QGz$ is the vector of the initial quantum numbers, that is $\QGz = 
 (2,2,0,0,0)$ in a pp collision and $\QGz = (0,0,0,0,0)$ in a \ppb 
 collision.\\ 
 The invariant partition function of a single fireball is, by definition:

 \begin{equation}
     Z_i(\QGzi) = \sum_{\rm{states}} \, \E^{- \beta_i \cdot P_i} 
     \delta_{\QGi,\QGzi} \; ,
 \end{equation}
 where $P_i$ is its total four-momentum. The factor $\delta_{\QGi,\QGzi}$ is the 
 usual Kronecker tensor, which forces the sum to be performed only over the fireball 
 states whose quantum numbers $\QGi$ are equal to the particular set $\QGzi$. 
 It is worth emphasizing that this partition function corresponds to the {\em canonical}
 ensemble of statistical mechanics since only the states fulfilling a fixed chemical 
 requirement, as expressed by the factor $\delta_{\QGi,\QGzi}$, are involved in the 
 sum (1).\\
 By using the integral representation of $\delta_{\QGi,\QGzi}$:

 \begin{equation}
    \delta_{\QGi,\QGzi} = \frac{1}{(2\pi)^5} \int_{0}^{2\pi} \!\!\!\! \int_{0}^{2\pi}
       \!\!\!\! \int_{0}^{2\pi} \!\!\!\! \int_{0}^{2\pi} \!\!\!\! \int_{0}^{2\pi}  
        \!\! \dint^5 \phi \,\, \E^{\,\I\,(\QGzi - \QGi) \cdot \phi} \; ,
 \end{equation}
 Eq.~(1) becomes:

 \begin{equation}
     Z_i(\QGzi) = \sum_{\rm{states}} \!\! \frac{1}{(2\pi)^5} \int_{0}^{2\pi} 
      \!\!\!\! \ldots \int_{0}^{2\pi} \!\!\! \dint^5 \phi \,\, \E^{- \beta_i \cdot P_i} 
      \E^{\,\I\, (\QGzi -\QGi) \cdot \phi} .
 \end{equation} 
 This equation could also have been derived from the general expression of partition 
 function of systems with internal symmetry \cite{zf1,zf2} by requiring a U(1)$^5$ 
 symmetry group, each U(1) corresponding to a conserved quantum number; that was the 
 procedure taken in ref. \cite{beca}.\\ 
 The sum over states in Eq.~3 can be worked out quite 
 straightforwardly for a hadron gas of $N_B$ boson species and $N_F$ 
 fermion species. A state is specified by a set of occupation numbers 
 $\{n_{j,k}\}$ for each phase space cell $k$ and for each particle 
 species $j$. Since $P_i=\sum_{j,k} p_{k} n_{j,k}$ and $\QGi= 
 \sum_{j,k} \qj n_{j,k}$, where $\qj = (Q_j,N_j,S_j,C_j,B_j)$ is the 
 quantum numbers vector associated to the $j^{th}$ particle species, 
 the partition function (3) reads, after summing over states:  

 \begin{equation}
  Z_i(\QGzi) = \frac{1}{(2\pi)^5} \int \, \dint^5 \phi \,\, 
      \E^{\,\I\, \QGzi \cdot \phi} \exp [ {\sum_{j=1}^{N_B} \sum_k \, \log \, 
       (1 - \E^{-\beta_i \cdot p_{k}-\I \qj \cdot \phi})^{-1}} 
  + {\sum_{j=1}^{N_F} \sum_k \, \log \, (1+\E^{-\beta \cdot p_{k} 
         -\I \qj \cdot \phi})}] \; .
 \end{equation}
 The last expression of the partition function is manifestly Lorentz-invariant
 because the sum over phase space is a Lorentz-invariant 
 operation which can be performed in any frame. The most suitable one 
 is the fireball rest frame, where the four-vector $\beta_i$ reduces 
 to: 

 \begin{equation}
 \beta_i= (\frac{1}{T_i},0,0,0)
 \end{equation}
 $T_i$ being the temperature of the fireball. Moreover, the sum over phase space cells 
 in Eq.~(4) can be turned into an integration over momentum space going to the 
 continuum limit:

 \begin{equation}
 \sum_{k} \longrightarrow (2J_j+1) \, \frac{V}{(2\pi)^3} \int \dint^3 p  \; ,
 \end{equation}  
 where $V$ is the fireball volume and $J_j$ the spin of the $j^{th}$ 
 hadron. As in previous studies on \ee collisions \cite{beca} and 
 heavy ions collisions \cite{raf}, we supplement the ordinary 
 statistical mechanics formalism with a strangeness suppression factor 
 $\gamma_s$ accounting for a partial strangeness phase space 
 saturation\footnote{Possible charm and beauty suppression parameters 
 $\gamma_c$ and $\gamma_b$ are unobservable, see also Appendix C.}; 
 actually the Boltzmann factor $\E^{-\beta \cdot p_{k}}$ of any hadron 
 species containing $s$ strange valence quarks or anti-quarks is 
 multiplied by $\gamma_s^{s}$. With the transformation (6) and 
 choosing the fireball rest frame to perform the integration, the sum 
 over phase space in Eq.~(4) becomes: 
 
 \begin{eqnarray}
  && \sum_{k} \, \log \, (1 \pm \gamma_s^{s_j} \E^{-\beta_i \cdot p_{k} -\I \qj
   \cdot \phi})^{\pm 1} \longrightarrow \nonumber \\
  && \frac{(2J_j+1)\,V_i}{(2\pi)^3} \! \int \dint^3 p \,\, \log \, (1 \pm \gamma_s^{s_j} 
      \E^{-\sqrt{p^2+m_j^2}/T_i -\I \qj \cdot \phi})^{\pm 1} \equiv V_i \, 
     F_j(T_i,\gamma_s,\PG) \; ,
 \end{eqnarray}
 where the upper sign is for fermions, the lower for bosons and $V_i$ 
 is the fireball volume in its rest frame; the function 
 $F_j(T_i,\gamma_s,\PG)$ is a shorthand notation of the momentum integral in 
 Eq.~(7). Hence, the partition function (4) can be written: 

 \begin{equation}
    Z_i(\QGzi) = \frac{1}{(2\pi)^5} \int \, \dint^5 \phi \,\, 
    \E^{\,\I\, \QGzi \cdot \phi} \exp \, [V_i \sum_j F_j(T_i,\gamma_s,\PG)] \; .
 \end{equation}   
 The mean number $\langle n_j \rangle_i$ of the $j^{th}$ particle 
 species in the $i^{th}$ fireball can be derived from $Z(\QGzi)$ by 
 multiplying the Boltzmann factor $\exp\,(-\sqrt{p^2+m_j^2}/T)$, in the 
 function $F_j$ in Eq.~(8) by a fictitious fugacity $\lambda_j$ and 
 taking the derivative of $\log Z_i(\QGzi,\lambda_j)$ with respect to 
 $\lambda_j$ at $\lambda_j = 1$: 
    
 \begin{equation}
   \langle n_j \rangle_i \,\, = \frac {\partial}{\partial \lambda_j} 
   \log Z_i(\QGzi,\lambda_j) \Big|_{\lambda_j=1}   \; .  
 \end{equation}
 The partition function $Z_i(\QGi,\lambda_j)$ supplemented with the 
 $\lambda_j$ factor is still a Lorentz-invariant quantity and so is 
 the mean number $\langle n_j \rangle_i$. From a more physical point of 
 view, this means that the average multiplicity of any hadron does not 
 depend on fireball collective motion, unlike its mean number in a 
 particular momentum state.\\        
 The overall average multiplicity of the $j^{th}$ hadron, for a set of $N$ fireballs in 
 a certain quantum configuration $\{\QG_1^0,\ldots,\QG_N^0\}$ is the sum 
 of all mean numbers of that hadron in each fireball:

 \begin{equation}
   \langle n_j \rangle \,\, 
   = \sum_{i=1}^N \frac {\partial}{\partial \lambda_j} 
   \log Z_i(\QGzi,\lambda_j) \Big|_{\lambda_j=1}  
   = \frac {\partial}{\partial \lambda_j} \log \prod_{i=1}^N 
     Z_i(\QGzi,\lambda_j) \Big|_{\lambda_j=1} \; .  
 \end{equation} 
 In general, as the quantum number configurations may fluctuate, 
 hadron production should be further averaged over all possible 
 fireballs configurations ${\QG_1^0,\ldots,\QG_N^0}$ fulfilling the 
 constraint $\sum_{i=1}^N \QGzi = \QGz$. To this end, suitable 
 weights $w(\QG_1^0,\ldots,\QG_N^0)$, representing the probability of 
 configuration $\{\QG_1^0,\ldots,\QG_N^0\}$ to occur for a set of $N$ 
 fireballs, must be introduced. Basic features of those weights are:

 \begin{eqnarray}
     && w(\QG_1^0,\ldots,\QG_N^0) = 0  \qquad{\rm{if}}\quad
          \sum_{i=1}^N \QGzi \neq \QGz \,, \nonumber \\
     && \sum_{\QG_1^0,\ldots,\QG_N^0} \!\!\!\! w(\QG_1^0,\ldots,\QG_N^0) = 1 \,.
 \end{eqnarray}
 For the overall average multiplicity of hadron $j$ we get:

 \begin{equation}
  \langle\!\langle n_j \rangle\!\rangle = \!\!\!\! 
     \sum_{\QG_1^0,\ldots,\QG_N^0} \!\!\!\! 
      w(\QG_1^0,\ldots,\QG_N^0) \frac {\partial}{\partial \lambda_j} \log 
     \prod_{i=1}^N Z_i(\QGzi,\lambda_j) \Big|_{\lambda_j=1}  \; .
 \end{equation}  
 There are infinitely many possible choices of the weights 
 $w(\QG_1^0,\ldots,\QG_N^0)$, all of them equally legitimate. However, 
 one of them is the most pertinent from the statistical mechanics point 
 of view, namely: 
 
 \begin{equation}
     w(\QG_1^0,\ldots,\QG_N^0) = \frac{\delta_{\zum_i \QGzi,\QGz} \prod_{i=1}^N 
      Z_i(\QGzi)}{\sum_{\QG_1^0,\ldots,\QG_N^0} \!\!\! \delta_{\zum_i \QGzi,\QGz} 
      \prod_{i=1}^N Z_i(\QGzi)}  .    
 \end{equation}
 It can be shown indeed that this choice corresponds to the minimal deviation 
 from statistical equilibrium of the system as a whole. In fact, putting weights (13) 
 in the Eq.~(12), one obtains:

 \begin{equation}  
 \langle\!\langle n_j \rangle\!\rangle = \frac {\partial}{\partial \lambda_j} 
        \log \!\!\!\! \sum_{\QG_1^0,\ldots,\QG_N^0}  \!\!\!\! 
        \delta_{\zum_i \QGzi,\QGz} \prod_{i=1}^N 
        Z_i(\QGzi,\lambda_j)\Big|_{\lambda_j=1}. \!\!\!\!  
 \end{equation}   
 This means that the average multiplicity of any hadron can be derived 
 from the following function of $\QGz$: 

 \begin{equation}
 Z(\QGz) = \!\!\!\! \sum_{\QG_1^0,\ldots,\QG_N^0} \!\!\!\! \delta_{\zum_i \QGzi,\QGz} 
                    \prod_{i=1}^N Z_i(\QGzi) \; , 
 \end{equation} 
 with the same recipe given for a single fireball in Eq.~(9). By using 
 expression (1) for the partition functions $Z_i(\QGzi)$, Eq.~(15) 
 becomes: 

 \begin{equation}
 Z(\QGz)= \!\!\!\! \sum_{\QG_1^0,\ldots,\QG_N^0} \!\!\!\! \delta_{\zum_i \QGzi,\QGz} 
  \prod_{i=1}^N \sum_{\rm{states}_i} \, \E^{- \beta_i \cdot P_i} \delta_{\QGzi,\QGi} \; .                   
 \end{equation} 
 Since

 \begin{equation}
  \sum_{\QG_1^0,\ldots,\QG_N^0} \!\!\!\! \delta_{\zum_i \QGzi,\QGz} \,\, 
   \delta_{\QGzi,\QGi} = \delta_{\zum_i \QGi,\QGz} \; , 
 \end{equation} 
 the function (16) can be written as

 \begin{equation}
  Z(\QGz)= \sum_{\rm{states}_1} \!\! \ldots \!\! 
  \sum_{\rm{states}_N} \, \E^{- \beta_1 \cdot P_1} \ldots \E^{- 
  \beta_N \cdot P_N} \delta_{\zum_i \QGi,\QGz} \; .                    
 \end{equation}  
 This expression demonstrates that $Z(\QGz)$ may be properly called 
 the {\em global partition function} of a system split into $N$ 
 subsystems which are in mutual chemical equilibrium but not in mutual 
 thermal and mechanical equilibrium. Indeed it is a Lorentz-invariant 
 quantity and, in case of complete equilibrium, i.e.  
 $\beta_1=\beta_2=\ldots=\beta_N\equiv \beta$, it would reduce to:

 \begin{equation}
  Z(\QGz) =\sum_{\rm{states}_1} \!\! \ldots \!\! \sum_{\rm{states}_N}
  \, \E^{- \beta \cdot (P_1 + \ldots \cdot P_N)} \delta_{\zum_i \QGi,\QGz} 
   = \sum_{\rm{states}} \, \E^{- \beta \cdot P} \delta_{\QG,\QGz}  \; ,                  
 \end{equation}    
 which is the basic definition of the partition function.\\
 To summarize, the choice of weights (13) allows the construction of a 
 system which is out of equilibrium only by virtue of its subdivision 
 into several parts having different temperatures and velocities.
 Another very important consequence of that choice is the following: 
 if we assume that the freeze-out temperature of the various fireballs 
 is constant, that is $T_1 = \ldots = T_N \equiv T$, and that the 
 strangeness suppression factor $\gamma_s$ is constant too, then the 
 global partition function (18) has the following expression: 

 \begin{equation}
    Z(\QGz) = \frac{1}{(2\pi)^5} \int \, \dint^5 \phi \,\, 
    \E^{\,\I\, \QGz \cdot \phi} \exp \, [ (\zum_i V_i) \sum_j 
    F_j(T,\gamma_s,\PG)]. 
 \end{equation}
 Here the $V_i$'s are the fireball volumes in their own rest frames; a 
 proof of (20) \cite{tesi} is given in Appendix A. Eq.~(20) demonstrates that 
 the global partition function has the same functional form (3), (4), 
 (8) as the partition function of a single fireball, once the volume $V_i$ 
 is replaced by the {\em global volume} $V \equiv \sum_{i=1}^N V_i$.
 Note that the global volume absorbs any dependence of the global 
 partition function (20) on the number of fireballs $N$. Thus, 
 possible variations of the number $N$ and the size $V_i$ of fireballs 
 on an event by event basis can be turned into fluctuations of the 
 global volume. In the remainder of this Section and in Sects.~3, 4 we will 
 ignore these fluctuations; in Sect. 5 it will be shown that they do 
 not affect any of the following results on the average hadron
 multiplicities.\\    
 The average multiplicity of the $j^{th}$ hadron can be determined 
 with the formulae (14)-(15), by using expression (20) for the function 
 $Z(\QGz)$: 
 
 \begin{eqnarray}
   \langle\!\langle n_j \rangle\!\rangle &=&  \frac{1}{(2\pi)^5}
    \int \dint^5 \phi \,\, \E^{\,\I\, \QGz \cdot \phi} \exp  
    [ V \sum_j F_j(T,\gamma_s,\PG)] 
    \nonumber \\ 
   &\times& \frac {(2J_j+1)\,V}{(2\pi)^3} \int 
      \frac {\dint^3 p}{\gamma_s^{-s_j}
      \exp \,(\sqrt{p^2+m^2_j}/T+\I \qj \cdot \PG) \pm 1} \; , 
 \end{eqnarray}   
 where the upper sign is for fermions and the lower for bosons. This 
 formula can be written in a more compact form as a series: 

 \begin{equation}  
   \langle\!\langle n_j \rangle\!\rangle =  
    \sum_{n=1}^{\infty} (\mp 1)^{n+1} \,\gamma_s^{n s_j} 
    z_{j(n)} \, \frac{Z(\QGz-n\qj)}{Z(\QGz)} \; ,
 \end{equation}
 where the functions $z_{j(n)}$ are defined as:

 \begin{equation}  
    z_{j(n)} \equiv (2J_j+1)\, \frac{V}{(2\pi)^3} \int \dint^3 
   p \, \exp \, (-n \sqrt{p^2+m^2_j}/T) = 
   (2J_j+1) \, \frac{VT}{2\pi^2 n} \, m_j^2 \, 
   {\rm{K}}_2(\frac{nm_j}{T}) \; .  
 \end{equation}
 K$_2$ is the McDonald function of order 2. Eq.~(22) is the final 
 expression for the average multiplicity of hadrons at freeze-out.
 Accordingly, the production rate of a hadron species depends only 
 on its spin, mass, quantum numbers and strange quark content.\\ 
 The {\em chemical factors} $Z(\QGz-n\qj)/Z(\QGz)$ in Eq.~(22) are a 
 typical feature of the canonical approach due to the requirement of exact 
 conservation of the initial set of quantum numbers. These factors 
 suppress or enhance production of particles according to the 
 vicinity of their quantum numbers to the initial $\QGz$ vector. The 
 behaviour of $Z(\QG)$ as a function of electric charge, baryon number 
 and strangeness for suitable $T$, $V$ and $\gamma_s$ values is shown 
 in Fig. 1; for instance, it is evident that the baryon chemical 
 factors $Z(0,N,0,0,0)/Z(0,0,0,0,0)$ connected with an initially 
 neutral system play a major role in determining the baryon
 multiplicities. The ultimate physical reason of ``charged" particle
 ($\qj \neq 0$) suppression with respect to ``neutral" ones ($\qj = 
 0$), in a completely neutral system ($\QGz = 0$), is the necessity, 
 once a ``charged" particle is created, of a simultaneous creation of 
 an anti-charged particle in order to fulfill the conservation laws. 
 In a {\em finite system} this pair creation mechanism is the more 
 unlikely the more massive is the lightest particle needed to 
 compensate the first particle's quantum numbers. For instance, once a 
 baryon is created, at least one anti-nucleon must be generated, which 
 is rather unlikely since its mass is much greater than the temperature 
 and the total energy is finite. On the other hand, if a non-strange 
 charged meson is generated, just a pion is needed to balance the 
 total electric charge; its creation is clearly a less unlikely event 
 with respect to the creation of a baryon as the energy to be spent is 
 lower. This argument illustrates why the dependence of $Z(\QG)$ on 
 the electric charge is much milder that on baryon number and 
 strangeness (see Fig. 1). In view of that, the dependence of 
 $Z(\QG)$ on electric charge was neglected in the previous study on 
 hadron production in \ee collisions \cite{beca}. 
 These chemical suppression effects are not accountable in a 
 grand-canonical framework; in fact, in a completely neutral system, 
 all chemical potentials should be set to zero and consequently 
 ``charged" particles do not undergo any suppression with respect to 
 ``neutral" ones.\\
 A compact analytic expression for the function $Z(\QG)$ does not exist. 
 However, an approximation of $Z(\QG)$ valid for large global volumes 
 (see Appendix B) exists in which chemical factors reduce to a product 
 of a chemical-potential-like factor and an additional multivariate gaussian 
 factor having no correspondence in the grand-canonical framework. The 
 gaussian factor tends to 1 for $V \rightarrow \infty$ proving the 
 equivalence between canonical and grand-canonical approaches for 
 large systems.\\ 
 The global partition function (18) has to be further modified in \ppb 
 collisions owing to a major effect in such reactions, the {\it leading 
 baryon effect} \cite{zichi}. Indeed, the sum (18) includes states 
 with vanishing net absolute value of baryon number, whereas in \ppb 
 collisions at least one baryon-antibaryon pair is always observed. 
 Hence, the simplest way to account for the leading baryon effect is to 
 exclude those states from the sum. Thus, if $|N| = \sum_i |N_i|$ 
 denotes the absolute value of the baryon number of the system, the 
 global partition function (18) should be turned into: 

 \begin{equation}
   Z = \!\!\!\! \sum_{\rm{states}_1} \!\! \ldots \!\! 
  \sum_{\rm{states}_N} \, \E^{- \beta_1 \cdot P_1} \ldots \E^{- 
  \beta_N \cdot P_N} \delta_{\zum_i \QGi,\QGz} 
  - \sum_{\rm{states}_1} \!\! \ldots \!\! 
  \sum_{\rm{states}_N} \, \E^{- \beta_1 \cdot P_1} \ldots \E^{- 
  \beta_N \cdot P_N} \delta_{\zum_i \QGi,\QGz} \delta_{|N|, 0}   \; .           
 \end{equation}
 The first term, that we define as $Z_1(\QGz)$, is equal to the 
 function $Z(\QGz)$ in Eqs.~(18), (20), while the second term is the 
 sum over all states having vanishing net absolute value of baryon 
 number. The absolute value of baryon number can be treated as a new 
 independent quantum number so that the processing of the partition 
 function described in Eqs.~(1)-(3) can be repeated for the second 
 term in Eq.~(24) with a U(1)$^6$ symmetry group. Accordingly, this 
 term can be naturally denoted by $Z_2(\QGz,0)$, so that Eq.~(24) 
 reads: 
 
 \begin{equation}
  Z = Z_1(\QGz) - Z_2(\QGz,0)  \; .
 \end{equation}
 By using the integral representation of $\delta_{|N|, 0}$ 

 \begin{equation}
    \delta_{|N|, 0} = \frac{1}{2\pi} \int_{0}^{2\pi} \!\! \dint \psi 
      \,\, \E^{\,\I\, |N| \cdot \psi}
 \end{equation}
 in the second term of Eq.~(24), one gets:

 \begin{eqnarray}
  Z_2(\QGz,0) &=& \frac{1}{(2\pi)^6} \int  \dint^5 \phi \,\, 
       \E^{\,\I\, \QGz \cdot \phi} \exp [V \sum_j F_j(T,\gamma_s,\PG)] \nonumber \\ 
          &\times& \int \dint \psi \,\, \exp [\sum_j 
    \frac{(2J_j+1)V}{(2\pi)^3}  \int \dint^3 p \, \log \, (1 + \gamma_s^{s_j} 
      \E^{-\sqrt{p^2+m_j^2}/T -\I \qj \cdot \phi - \,\I\, \psi})] \;, 
 \end{eqnarray}
 where the first sum over $j$ runs over all mesons and the second over 
 all baryons. The average multiplicity of any hadron species can be 
 derived from the global partition function (25) with the usual 
 prescription: 

 \begin{equation}
   \langle\!\langle n_j \rangle\!\rangle = \frac {\partial}{\partial \lambda_j} 
         \log Z(\lambda_j) \Big|_{\lambda_j=1}  \; .
 \end{equation}
  
\section{Fit procedure and data set}

 The model described so far has three free parameters: the temperature 
 $T$, the global volume $V$ and the strangeness suppression parameter 
 $\gamma_s$. They will be determined by a fit to the available data 
 on hadron inclusive production at each centre of mass energy.
 Eq.~(22) yields the mean number of hadrons emerging directly from the 
 thermal source at freeze-out, the so-called primary hadrons
 \cite{beca,giova}, as a function of the three free parameters. After 
 freeze-out, primary hadrons trigger a decay chain process which must 
 be properly taken into account in a comparison between model 
 predictions and experimental data, as the latter generally embodies 
 both primary hadrons and hadrons generated by heavier particles 
 decays. Therefore, in order to calculate overall average 
 multiplicities to be compared with experimental data, the primary 
 yield of each hadron species, determined according to Eq.~(22) (or 
 (28) for \ppb collisions) is added to the contribution stemming from 
 the decay of heavier hadrons, which is calculated by using 
 experimentally known decay modes and branching ratios 
 \cite{pdg,jet}.\\ 
 The calculation of the average multiplicity of primaries according to 
 Eq.~(22) involves several rather complicated five-dimensional 
 integrals which have been calculated numerically after some useful 
 approximations, described in the following. Since the temperature is 
 expected to be below 200 MeV, the primary production rate of all 
 hadrons, except pions, is very well approximated by the first term of 
 the series (22): 

 \begin{equation}  
   \langle\!\langle n_j \rangle\!\rangle \simeq  
   \gamma_s^{s_j}\, z_j \, \frac{Z(\QGz-\qj)}{Z(\QGz)} \; ,
 \end{equation}  
 where we have put $z_j \equiv z_{j(1)}$. This approximation corresponds 
 to the Boltzmann limit of Fermi and Bose statistics. Actually, for a 
 temperature of 170 MeV, the primary production rate of K$^+$, the 
 lightest hadron after pions, differs at most (i.e. without the
 strangeness suppression parameter and the chemical factors which further 
 reduce the contribution of neglected terms) by 1.5\% from that 
 calculated with Eq.~(29), well within usual experimental 
 uncertainties. Corresponding Boltzmannian approximations can be made 
 in the function $Z(\QG)$, namely

 \begin{equation}
  \log \, (1\pm \E^{-\sqrt{p^2+m^2_j}/T -\I 
    \qj \cdot \phi })^{\pm 1} \simeq \E^{-\sqrt{p^2+m^2_j}/T -\I 
    \qj \cdot \phi } \; ,
 \end{equation}
 which turns Eq.~(20) (for a generic $\QG$) into:

 \begin{equation}
      Z(\QG) \simeq \frac{1}{(2\pi)^5} \int \, \dint^5 \phi \,\, 
       \E^{\,\I\, \QG \cdot \phi} \exp \, [ \sum_j z_j \gamma_s^{s_j} 
       \E^{-\I \qj \cdot \phi} + \sum_{j=1}^{3} \frac{V}{(2\pi)^3} \int \dint^3 p \, 
      \log \, (1 - \E^{-\sqrt{p^2+m_j^2}/T -\I \qj \cdot \phi})^{-1}] \; ,
 \end{equation} 
 where the first sum runs over all hadrons except pions and the 
 second over pions.\\
 As a further consequence of the expected temperature value, the $z$ 
 functions of all charmed and bottomed hadrons are very small: with 
 $T= 170$ MeV and a primary production rate of K mesons of the order 
 of one, as the data states, the $z$ function of the lightest charmed 
 hadron, D$^0$, turns out to be $\approx 10^{-4}$; chemical factors 
 produce a further suppression of a factor $\approx 10^{-4}$. 
 Therefore, thermal production of heavy flavoured hadrons can be 
 neglected, as well as their $z$ functions in the exponentiated sum in 
 Eq.~(31), so that the integration over the variables $\phi_4$ and 
 $\phi_5$ can be performed: 

 \begin{eqnarray}
    Z(\QG,C,B) &\simeq& \frac{1}{(2\pi)^3} \int \, \dint^3 \phi 
   \,\, \E^{\,\I\, \QG \cdot \phi} \exp \, [\sum_j z_j \gamma_s^{s_j} 
   \E^{-\I \qj \cdot \phi} 
    \nonumber \\
     &-& \sum_{j=1}^{3} \frac{V}{(2\pi)^3} \int \dint^3 p \, 
      \log \, (1 - \E^{-\sqrt{p^2+m_j^2}/T -\I \qj \cdot \phi})] 
      \, \delta_{C,0} \delta_{B,0} \equiv \zeta(\QG) \, \delta_{C,0} \delta_{B,0} \; .
 \end{eqnarray} 
 $\QG$ and $\qj$ are now three-dimensional vectors consisting of
 electric charge, baryon number, and strangeness; the five-dimensional 
 integrals have been reduced to three-dimensional ones.\\ 
 Apart from the hadronization contribution, which is expected to be 
 negligible in this model, production of heavy flavoured hadrons in 
 hadronic collisions mainly proceeds from hard perturbative QCD 
 processes of c${\rm{\bar{c}}}$ and b${\rm{\bar{b}}}$ pairs 
 creation. The fact that promptly generated heavy quarks do not 
 reannihilate into light quarks indicates a strong deviation from 
 statistical equilibrium of charm and beauty, much stronger than
 the strangeness suppression linked with $\gamma_s$. Nevertheless, it 
 has been found in \ee collisions \cite{beca} that the relative 
 abundances of charmed and bottomed hadrons are in agreement with 
 those predicted by the statistical equilibrium assumption, confirming 
 its full validity for light quarks and quantum numbers associated to 
 them. The additional source of heavy flavoured hadrons arising from 
 perturbative processes can be accounted for by modifying the partition 
 function (31). In particular, the presence of one heavy flavoured 
 hadron and one anti-flavoured hadron should be demanded in a fraction 
 of events $f= \sigma ({\rm{pp(\bar p)}} \rightarrow \cc) / \sigma 
 ({\rm{pp(\bar p)}})$ (or $f= \sigma ({\rm{pp(\bar p)}} 
 \rightarrow \bb) / \sigma ({\rm{pp(\bar p)}})$) where 
 $\sigma({\rm{pp(\bar p)}})$ is meant to be the total inelastic or 
 non-single-diffractive cross section. Accordingly, the partition 
 function to be used in events with a perturbative 
 c${\rm{\bar{c}}}$ pair, is, by analogy with Eq.~(24)-(25) and the
 leading baryon effect: 

 \begin{eqnarray}
  Z &=& \!\!\!\! \sum_{\rm{states}_1} \!\! \ldots \!\! \sum_{\rm{states}_N}
  \, \E^{- \beta_1 \cdot P_1} \ldots \E^{- \beta_N \cdot P_N} \delta_{\zum_i \QGi,\QGz} 
  \nonumber \\
  &-& \sum_{\rm{states}_1} \!\! \ldots \!\! \sum_{\rm{states}_N}
  \, \E^{- \beta_1 \cdot P_1} \ldots \E^{- \beta_N \cdot P_N} \delta_{\zum_i \QGi,\QGz}
  \delta_{|C|, 0} \equiv Z_1(\QGz) - Z_2(\QGz,0)  \: ,            
 \end{eqnarray} 
 where $|C|$ is the absolute value of charm. The primary yield of 
 charmed hadrons, calculated according to Eq.~(28) and partition 
 function (33), is derived in Appendix C.\\ 
 A significant production rate of heavy flavoured hadrons might affect 
 light hadrons abundances through decay feed-down, so it is important 
 to know how large the fraction $f$ is. Available data on charm 
 cross-sections \cite{charm} indicate a fraction $f \approx 10^{-2} 
 \div 10^{-3}$ at centre of mass energies $< 30$ GeV and, 
 consequently, much lower values for bottom quark production. 
 Therefore, the perturbative production of heavy quarks can be 
 neglected as long as one deals with light flavoured hadron production 
 at $\sqrt s < 30$ GeV. We assume that it may be neglected at any 
 centre of mass energy; this point will be discussed in more detail in 
 the next section.
\begin{scriptsize}
\begin{table*}
\caption[]{\footnotesize{Values of fitted parameters in pp and \ppb collisions. The 
  normalization parameter $V T^3$ is better suited than $V$ in the fit 
  because is less correlated to the temperature. The additional errors 
  within brackets have been estimated by excluding some data points 
  and repeating the fit. Also quoted is the correlation parameter 
  between $T$ and $VT^3$.}} 
\begin{tabular}{llllll}
\hline\noalign{\smallskip} 
   $\sqrt s$ (GeV)          & $T$ (MeV)             & $V T^3$                  &  $\gamma_s$                & $\chi^2/$dof  &  $\rho(T, VT^3)$\\ 
\noalign{\smallskip}\hline
\noalign{\smallskip}
  pp collisions             & & & &  \\  
\noalign{\smallskip}\hline
\noalign{\smallskip}
 $19.4 \div 19.6$           &$190.8\pm27.4$         & $5.79\pm3.05$            & $0.463\pm0.037$            &  6.38/4       &  -0.999   \\
   23.8                     &$194.4\pm17.3$         & $6.34\pm2.49$            & $0.460\pm0.067$            &  2.43/2       &  -0.936   \\  
   26.0                     &$159.0\pm9.5$          & $13.36\pm2.66$           & $0.570\pm0.030$            &  1.86/2       &  -0.993   \\ 
 $27.4 \div 27.6$           &$169.0\pm2.1\;(\pm3.4)$& $11.04\pm0.69\;(\pm1.4)$ & $0.510\pm0.011\;(\pm0.025)$&  136.4/27     &  -0.972   \\ 
\noalign{\smallskip}\hline
\noalign{\smallskip}
   \ppb collisions          & & & &  \\         
\noalign{\smallskip}\hline
\noalign{\smallskip}
  200                       &$175.4\pm14.8$         & $24.26\pm7.89$           & $0.537\pm0.066$            &  0.698/2      &  -0.989  \\ 
  546                       &$181.7\pm17.7$         & $28.5\pm10.4$            & $0.557\pm0.052$            &  3.80/1       &  -0.993  \\ 
  900                       &$170.2\pm11.8$         & $43.2\pm11.8$            & $0.578\pm0.063$            &  1.79/2       &  -0.982  \\ 
\noalign{\smallskip}\hline 
\end{tabular}
\end{table*}
\end{scriptsize}
 \\ 
 All light flavoured hadrons and resonances with a mass $< 1.7$ GeV 
 have been included among the primary generated hadron species; the 
 effect of this cut-off on obtained results will be discussed in the 
 next section. The mass of resonances with $\Gamma > 1$ MeV has been 
 distributed according to a relativistic Breit-Wigner function within 
 $\pm 2\Gamma$ from the central value. The $\gamma_s$ strangeness 
 suppression factor has also been applied to neutral mesons such as 
 $\phi$, $\omega$, etc. according to the their strange valence quark
 content; mixing angles quoted in ref. \cite{pdg} have been used. Once the
 average multiplicities of the primary hadrons have been calculated as 
 a function of the three parameters $T$, $V$ and $\gamma_s$, the decay 
 chain is performed until $\pi$, $\mu$, K$^{\pm}$, K$^0$, $\Lambda$, 
 $\Xi$, $\Sigma^{\pm}$, $\Omega^-$ or stable particles are reached, in 
 order to match the average multiplicity definition in pp and \ppb 
 collisions experiments. It is worth mentioning that, unlike pp and 
 \ppb, all \ee colliders experiments also include the decay products
 of K$^0_s$, $\Lambda$, $\Xi$, $\Sigma^{\pm}$ and $\Omega^-$ in their 
 multiplicity definition.\\ 
 Finally, the overall yield is compared with experimental 
 measurements, and the $\chi^2$: 

 \begin{equation} 
  \chi^2 = \sum_i ({\rm{theo}}_i-{\rm{expe}}_i)^2/{\rm{error}}_i^2 
 \end{equation}
 is minimized.\\
 As far as the data set is concerned, we used all available 
 measurements of hadron multiplicities in non-single-diffractive \ppb 
 and inelastic pp collisions down to a centre of mass energy of about 
 19 GeV (see Tables 2 and 3), fulfilling the following quality 
 requirements: 
 \begin{enumerate}
  \item the data is the result of an actual experimental measurement 
        and not a derivation based on isospin symmetry arguments; 
        indeed, this model predicts slight violations of isospin 
        symmetry due to mass differences;  
  \item the multiplicity definition is unambiguous, that means it is 
        clear what decay products are included in the quoted numbers; 
        actually, all referenced papers take the multiplicity 
        definition previously mentioned; 
  \item the data is the result of an extrapolation of a spectrum 
        measured over a large kinematical region.  
 \end{enumerate} 
 Some referenced papers about pp collisions quote cross sections 
 instead of average multiplicities. In some cases (e.g. ref. 
 \cite{agui}) both of them are quoted for some particles, which makes 
 it possible to obtain the average multiplicity of particles for which 
 only the cross section is given. Otherwise, total inelastic pp cross 
 sections have been extracted from other papers.\\
 Whenever several measurements at the same centre of mass energy have been 
 available, averages have been calculated according to a weighting 
 procedure described in ref. \cite{dean} prescribing rescaling of 
 errors to take into account {\em a posteriori} correlations and 
 disagreements of experimental results.
\begin{scriptsize}
\begin{table*}
\caption[]{\footnotesize{Measured average multiplicities compared with fitted 
 theoretical values in pp collisions. The first quoted error beside 
 measured values is the experimental error, the number within brackets 
 is the error due to uncertainty on masses, widths and branching 
 ratios of the various hadrons. Also quoted are the theoretical estimates
 of the fraction of primaries.}} 
\begin{tabular}{lllll}
\hline\noalign{\smallskip} 
    Particles          &  Measurement                  & Calculated        &  Primary fraction &  References \\ 
\noalign{\smallskip}\hline
\noalign{\smallskip}
$\sqrt s =19.4 \div 19.7$GeV&                          &                   &                   &               \\  
\noalign{\smallskip}\hline
\noalign{\smallskip}
   Neg. charged        & $2.85\pm0.040\;(\pm 0.063)$   &    2.798          &                   & \cite{all},\cite{bar} \\
     Charged           & $7.69\pm0.070\;(\pm0.13)$     &    7.620          &                   & \cite{all},\cite{bar} \\
      $\pi^0$          & $3.34\pm0.24\;(\pm0.11)$      &    3.404          & 0.173             & \cite{jaeg}$^{a}$     \\
      K$^0_s$          & $0.174\pm 0.013\;(\pm0.002 )$ &    0.160          & 0.325 - 0.369$^{b}$& \cite{jaeg},\cite{all}\\
      $\rho^0$         & $0.33 \pm0.06 \;(\pm 0.025 )$ &    0.448          & 0.568             & \cite{sing}           \\                      
      $\Lambda$        & $0.0977\pm0.0097\;(\pm0.0056)$&    0.110          & 0.243             & \cite{jaeg},\cite{all}\\
      $\bar\Lambda$    & $0.0136\pm0.0041\;(\pm0.0007)$&    0.0135         & 0.238             & \cite{jaeg},\cite{all}\\
\noalign{\smallskip}\hline
\noalign{\smallskip}
 $\sqrt s= 23.8 $ GeV  &                               &                   &                   &             \\ 
\noalign{\smallskip}\hline
\noalign{\smallskip}
     $\pi^0$           & $3.42\pm0.62\;(\pm 0.12)$     &    3.908          & 0.166             & \cite{kaf}$^{c}$ \\
      K$^0_s$          & $0.22\pm0.025\;(\pm0.003)$    &    0.198          & 0.319 - 0.362$^{b}$& \cite{lopi}      \\
 K$^{*+}$+K$^{*-}$     & $0.137\pm0.043\;(\pm0.002)$   &    0.165          & 0.658 - 0.492$^{d}$& \cite{lopi}$^{c}$\\
      $\Lambda$        & $0.11\pm 0.02\;(\pm0.007)$    &    0.126          & 0.238             & \cite{lopi}      \\
      $\bar\Lambda$    & $0.021\pm0.004\;(\pm0.001)$   &    0.0202         & 0.233             & \cite{lopi}      \\
\noalign{\smallskip}\hline
\noalign{\smallskip}
$\sqrt s = 26.0$ GeV   &                               &                   &                   &             \\ 
\noalign{\smallskip}\hline
\noalign{\smallskip}
   Neg. charged        & $3.53\pm0.05\;(\pm0.094)$     &    3.545          &                   & \cite{bail} \\
     Charged           & $9.06\pm 0.09\;(\pm0.18 )$    &    9.087          &                   & \cite{bail} \\
      K$^0_s$          & $0.26\pm0.01\;(\pm0.005)$     &    0.256          & 0.507 - 0.559$^{b}$& \cite{asai} \\
      $\Lambda$        & $0.12\pm 0.02\;(\pm0.009)$    &    0.147          & 0.295             & \cite{asai} \\
      $\bar\Lambda$    & $0.013\pm 0.004\;(\pm0.0007)$ &    0.0120         & 0.292             & \cite{asai} \\
\noalign{\smallskip}\hline
\noalign{\smallskip}
$\sqrt s=27.4 \div 27.6$ GeV &                               &                   &                   &             \\  
\noalign{\smallskip}\hline
\noalign{\smallskip}
    $\pi^+$            & $4.10\pm0.11\;(\pm0.15)$      &    4.147          & 0.293             & \cite{agui}\\
    $\pi^0$            & $3.87\pm0.12\;(\pm0.16)$      &    4.197          & 0.258             & \cite{agui}\\
    $\pi^-$            & $3.34\pm0.08\;(\pm0.12)$      &    3.269          & 0.223             & \cite{agui}\\
    K$^+$              & $0.331\pm0.016\;(\pm0.007)$   &    0.302          & 0.484             & \cite{agui}\\
    K$^-$              & $0.224\pm0.011\;(\pm0.004)$   &    0.182          & 0.380             & \cite{agui}\\
    K$^0_s$            & $0.232\pm0.011\;(\pm0.004)$   &    0.232          & 0.446 - 0.495$^{b}$& \cite{kich}\\
    $\eta$             & $0.30\pm0.02\;(\pm0.054)$     &    0.366          & 0.453             & \cite{agui}\\
    $\rho^0$           & $0.385\pm0.018\;(\pm0.038)$   &    0.543          & 0.628             & \cite{agui}\\
    $\rho^+$           & $0.552\pm0.083\;(\pm0.046)$   &    0.601          & 0.657             & \cite{agui}\\
    $\rho^-$           & $0.355\pm0.058\;(\pm0.033)$   &    0.421          & 0.569             & \cite{agui}\\
    $\omega$           & $0.390\pm0.024\;(\pm0.002)$   &    0.443          & 0.665             & \cite{agui}\\
    K$^{*+}$           & $0.132\pm0.016\;(\pm0.002)$   &    0.111          & 0.742             & \cite{agui}\\
    K$^{*-}$           & $0.088\pm0.012\;(\pm0.001)$   &    0.0617         & 0.628             & \cite{agui}\\
    K$^{*0}$           & $0.119\pm0.021\;(\pm0.002)$   &    0.0927         & 0.679             & \cite{agui}\\
$\bar{\rm{K}}^{*0}$& $0.0903\pm0.016\;(\pm0.001)$  &    0.0708         & 0.687             & \cite{agui}\\
    $\phi$             & $0.019\pm0.0018\;(\pm0.)$     &    0.0262         & 1.00              & \cite{agui}\\  
    f$_2(1270)$        & $0.092\pm0.012\;(\pm0.002)$   &    0.0684         & 0.845             & \cite{agui}\\
      p                & $1.20\pm0.097\;(\pm0.022)$    &    1.060          & 0.337             & \cite{agui}\\
 $\bar{\rm{p}}$    & $0.063\pm0.002\;(\pm0.001)$   &    0.0610         & 0.283             & \cite{agui}\\
   $\Lambda$           & $0.125\pm0.008\;(\pm0.008)$   &    0.136          & 0.276             & \cite{kich}\\
  $\bar\Lambda$        & $0.020\pm0.004\;(\pm 0.0008)$ &    0.0147         & 0.273             & \cite{kich}\\   
 $\Sigma^{+}$          & $0.048\pm0.015\;(\pm0.004)$   &    0.0423         & 0.688             & \cite{agui}$^{e}$\\
 $\Sigma^{-}$          & $0.0128\pm0.0061\;(\pm0.0032)$&    0.0310         & 0.592             & \cite{agui}$^{e}$\\
 $\Delta^{++}$         & $0.218\pm0.0031\;(\pm0.013)$  &    0.250          & 0.758             & \cite{agui}\\
 $\Delta^{0}$          & $0.141\pm0.0098\;(\pm0.0089)$ &    0.212          & 0.714             & \cite{agui}\\
 $\bar\Delta^{++}$     & $0.013\pm0.0049\;(\pm0.00049)$&    0.0111         & 0.548             & \cite{agui}\\
 $\bar\Delta^{0}$      & $0.0336\pm0.008\;(\pm0.0006)$ &    0.0165         & 0.697             & \cite{agui}\\
 $\Sigma^{*+}$         & $0.020\pm0.0025\;(\pm0.0011)$ &    0.0230         & 1.00              & \cite{agui}\\
 $\Sigma^{*-}$         & $0.010\pm0.0018\;(\pm0.0007)$ &    0.0139         & 1.00              & \cite{agui}\\
 $\Lambda(1520)$       & $0.017\pm0.0031\;(\pm0.0005)$ &    0.00996        & 1.00              & \cite{agui}\\
\noalign{\smallskip}\hline 
\noalign{\smallskip}
\multicolumn{5}{l}{$a$ - The $\pi^0$ multiplicity is defined in this paper as half 
                         the photon multiplicity; therefore, the experimental 
                         value}\\
\multicolumn{5}{l}{      has been fitted to half the number of photons coming 
                         not only from $\pi^0$ but also from $\eta$, $\omega$ and $\Sigma^0$ 
                         decays.}\\
\multicolumn{5}{l}{$b$ - Primary fraction of K$^0$ and $\bar{\rm{K}}^0$ respectively} \\
\multicolumn{5}{l}{$c$ - This paper quotes only the cross section. The multiplicity 
                         has been obtained by using the total inelastic cross}\\
\multicolumn{5}{l}{      section of 32.21 mb at $\sqrt s = 23.5$ GeV quoted in ref. \cite{ama}}\\
\multicolumn{5}{l}{$d$ - Primary fraction of K$^{*+}$ and K$^{*-}$ respectively} \\
\multicolumn{5}{l}{$e$ - Cross-reference to \cite{okuz}}\\
\noalign{\smallskip}
\noalign{\smallskip}\hline 
\end{tabular}
\end{table*}
\end{scriptsize}
 \\
 Since the decay chain is an essential step of the fitting procedure, 
 calculated theoretical multiplicities are affected by experimental 
 uncertainties on masses, widths and branching ratios of all involved 
 hadron species. In order to estimate the effect of these 
 uncertainties on the results of the fit, a two-step procedure for the fit 
 itself has been adopted: firstly, the fit has been performed with a 
 $\chi^2$ including only experimental errors and a set of parameters 
 $T_0$, $V_0$, $\gamma_{s0}$ has been obtained. Then, the various masses, 
 widths and branching ratios have been varied in turn by their errors, 
 as quoted in ref. \cite{pdg}, and new theoretical multiplicities 
 calculated, keeping the parameters $T_0$, $V_0$, $\gamma_{s0}$ fixed. 
 The differences between old and new theoretical multiplicity values 
 have been considered as additional systematic errors to be added in 
 quadrature to experimental errors. Finally, the fit has been 
 repeated with a $\chi^2$ including overall errors so as to obtain 
 final values for model parameters and for theoretical multiplicities.
 Among the mass, width and branching ratio uncertainties, only 
 those producing significant variations of final hadron yields 
 (actually more than 130) have been considered.
\begin{scriptsize} 
\begin{table*}
\caption[]{\footnotesize{Measured average multiplicities compared with fitted 
 theoretical values in \ppb collisions. The first quoted error beside 
 measured values is the experimental error, the number within brackets 
 is the error due to the uncertainty on masses, widths and branching 
 ratios of the various hadrons. Also quoted are the theoretical estimates 
 of the fraction of primaries.}} 
\begin{tabular}{lllll}
\hline\noalign{\smallskip} 
     Particle          &  Measurement                 & Calculated        &  Primary fraction &  References \\ 
\noalign{\smallskip}\hline
\noalign{\smallskip}                              
$\sqrt s =$ 200 GeV    &                              &                   &                   &             \\  
\noalign{\smallskip}\hline
\noalign{\smallskip}
     Charged           & $21.4\pm0.4\;(\pm0.72)$      &  21.27            &                   & \cite{anso}$^{a}$\\
      K$^0_s$          & $0.75\pm 0.09\;(\pm0.009)$   &  0.783            & 0.467             & \cite{anso} \\
      n                & $0.75\pm0.1\;(\pm0.05)$      &  0.794            & 0.291             & \cite{anso}$^{b}$\\                      
      $\Lambda$        & $0.23\pm0.06\;(\pm0.008)$    &  0.194            & 0.263             & \cite{anso} \\                                  
      $\Xi^-$          & $0.015\pm0.015\;(\pm0.0002)$ &  0.0123           & 0.579             & \cite{anso} \\ 
\noalign{\smallskip}\hline
\noalign{\smallskip}
$\sqrt s =$546 GeV     &                              &                   &                   &             \\ 
\noalign{\smallskip}\hline
\noalign{\smallskip}
     Charged           & $29.4\pm0.3\;(\pm0.96)$      &    29.25          &                   & \cite{anso}$^{a}$\\
      K$^0_s$          & $1.12\pm 0.08\;(\pm0.012 )$  &    1.139          &  0.441            & \cite{anso} \\
      $\Lambda$        & $0.265\pm0.055\;(\pm0.001)$  &    0.302          &  0.252            & \cite{anso} \\
      $\Xi^-$          & $0.05\pm0.015\;(\pm0)$       &    0.0228         &  0.567            & \cite{anso} \\
\noalign{\smallskip}\hline
\noalign{\smallskip}
$\sqrt s =$900 GeV     &                              &                   &                   &             \\ 
\noalign{\smallskip}\hline
\noalign{\smallskip}
     Charged           & $35.6\pm0.9\;(\pm1.2)$       &    35.15          &                   & \cite{anso}$^{a}$\\
      K$^0_s$          & $1.37\pm0.13\;(\pm0.02)$     &    1.437          & 0.497             & \cite{anso} \\
      n                & $1.0\pm0.2\;(\pm0.09)$       &    1.188          & 0.301             & \cite{anso}$^{b}$\\
      $\Lambda$        & $0.38\pm0.08\;(\pm0.01)$     &    0.323          & 0.269             & \cite{anso} \\
      $\Xi^-$          & $0.035\pm0.02\;(\pm0)$       &    0.0258         & 0.585             & \cite{anso} \\
\noalign{\smallskip}\hline
\noalign{\smallskip}
\multicolumn{5}{l}{$a$ - The charged track average multiplicity value quoted in this reference has been increased} \\
\multicolumn{5}{l}{      by one as leading particles, assumed to be one charged-neutral nucleon-antinucleon}\\
\multicolumn{5}{l}{      pair per event, were excluded.}\\
\multicolumn{5}{l}{$b$ - The neutron average multiplicity quoted in this reference has been increased}\\
\multicolumn{5}{l}{      by 0.5 as leading particles, assumed to be one charged-neutral nucleon-antinucleon}\\
\multicolumn{5}{l}{      pair per event, were excluded.}\\
\noalign{\smallskip}
\noalign{\smallskip}\hline
\end{tabular}
\end{table*}
\end{scriptsize}

\section{Results and checks} 

 The fitted values of the parameters $T$, $V$, $\gamma_s$ at 
 various centre of mass energy points are quoted in Table 1 while the 
 fitted values of average multiplicities are quoted in Table 2, 3 
 along with measured average multiplicities and the estimated primary 
 fraction. The fit quality is very good at almost all centre of mass 
 energies as demonstrated by the low values of $\chi^2$'s and by the 
 Figs. 2, 3, 4, 5, 6. Owing to the relatively large value of $\chi^2$ 
 at $\sqrt s = 27.4$ GeV, variations of fitted parameters larger than fit 
 errors must be expected when repeating the fit excluding data points 
 with the largest deviations from the theoretical values. Therefore, 
 the fit at $\sqrt s = 27.4$ GeV pp collisions has been repeated 
 excluding in turn ($\Delta^0$, $\rho^0$, $\phi$) and (K$^{-}$, 
 pions), respectively, from the data set; the maximum difference 
 between the new and old fit parameters has been considered as an 
 additional systematic error and is quoted in Table 1 within brackets.\\ 
 The fitted temperatures are compatible with a constant value at 
 freeze-out independently of collision energy and kind of 
 reaction (see Fig. 7). On the other hand, $\gamma_s$ exhibits a very slow rise 
 from 20 to 900 GeV (see Fig. 8); its value of $\simeq 0.5$ over 
 the whole explored centre of mass energy range proves that complete 
 strangeness equilibrium is not attained. Moreover, the temperature 
 value $\simeq 170$ MeV is in good agreement with that found in \ee 
 collisions \cite{beca,beca2} and in heavy ions collisions 
 \cite{satz}. On the other hand, the global volume does increase as a 
 function of centre of mass energy as it is proportional, for nearly 
 constant $T$ and $\gamma_s$, to overall multiplicity which indeed 
 increases with energy. Its values range from 6.4 fm$^3$ at $\sqrt s = 
 19.4$ GeV pp collisions, at a temperature of 191 MeV, up to 67 fm$^3$ 
 at $\sqrt s = 900$ GeV \ppb collisions at a temperature of 170 MeV.
 However, since volume values are strongly correlated to those of 
 temperature in the fit, errors turn out to be quite large and fit 
 convergence is slowed down; that is the reason why we actually fitted 
 the product $VT^3$ instead of $V$ alone.\\
 Once $T$, $V$ and $\gamma_s$ are determined by fitting average 
 multiplicities of some hadron species, their values can be used to 
 predict average multiplicities of any other species, at a given 
 centre of mass energy.\\
 Since the dependence of the chemical factors on the global volume $V$ 
 is quite mild in the region of interest (see Fig. 2), the hadron density
 mainly depends on 
 the temperature and $\gamma_s$ (cf. Eqs.~(22), (29)). Therefore, 
 constant values of temperature and $\gamma_s$ imply a nearly constant 
 hadron density at freeze-out, which turns out to be $\approx 0.4 \div 
 0.5$ hadrons/fm$^3$, as shown in Fig. 10, corresponding to a mean 
 distance between hadrons of approximately $\approx 1.6 \div 1.7$ fm.
 Unfortunately, due to its dramatic dependence on the temperature, all 
 density values, except that at $\sqrt s =27.4$ GeV, are affected by 
 large errors, and thus a definite claim of a constant freeze-out density
 cannot be made. The same statement is true for the pressure, also shown in 
 Fig. 10, whose definition is given in Appendix D.\\ 
 The physical significance of the results found so far depends 
 on their stability as a function of the various approximations and 
 assumptions which have been introduced. First, the temperature and 
 $\gamma_s$ values are low enough to justify the use of the Boltzmann 
 limits (29), (30) for all hadrons except pions, as explained in Sect.
 3. As far as the effect of a cut-off in the hadronic mass spectrum 
 goes, the most relevant test proving that our results so far do not 
 depend on it is the stability of the number of primary hadrons 
 against changes of the cut-off mass. The fit procedure intrinsically 
 attempts to reproduce fixed experimental multiplicities; if the 
 number of primary hadrons does not change significantly by repeating 
 the fit with a slightly lower cut-off, the production of heavier 
 hadrons excluded by the cut-off must be negligible, in particular 
 with regard to its decay contributions to light hadron yields. In 
 this spirit all fits have been repeated moving the mass cut-off value 
 from 1.7 down to 1.3 GeV in steps of 0.1 GeV, checking the stability 
 of the amount of primary hadrons as well as of the fit parameters. 
 It is worth remarking that the number of hadronic states with a mass 
 between 1.7 and 1.6 GeV is 238 out of 535 overall, so that their 
 exclusion is really a severe test for the reliability of the final 
 results. Figure 11 shows the model parameters and the primary hadrons 
 in \ppb collisions at $\sqrt s = 900$ GeV; above a cut-off of 1.5 GeV 
 the number of primary hadrons settles at an asymptotically stable 
 value, whilst the fitted values for $T$, $V$, $\gamma_s$ do not show 
 any particular dependence on the cut-off. Therefore, we conclude that 
 the chosen value of 1.7 GeV ensures that the obtained results are 
 meaningful.\\ 
 As mentioned in Sect. 3, the perturbative production of heavy 
 quarks has been neglected. This is legitimate in low energy pp 
 collisions, where it has been actually measured \cite{charm}, but
 not necessarily in $\sqrt s = {\cal{O}}(100)$ GeV \ppb collisions, 
 where no measurement exists and one has to rely on theoretical 
 estimates. In general, the latter predict very low b quark 
 cross sections, but a possibly not negligible c quark 
 production. We used the calculations of ref. \cite{vogt} 
 according to which the fraction $f$ of non-single-diffractive events 
 in which $\cc$ pairs are produced (see Sect. 3) rises as a function 
 centre of mass energy. We repeated the fit for \ppb collisions at 
 $\sqrt s = 900$, where the fraction $f$ is expected to be the 
 largest, by using the upper estimate of a cross section $\sigma 
 ({\rm{pp(\bar p)}}\rightarrow \cc) \simeq 12$ mb, corresponding to 
 $f \simeq 0.3$, in order to maximize the effect of charm production.
 The partition function to be used in such events is that in Eq.~(33) 
 with a further modification according to Eq.~(24) to take into 
 account the leading baryon effect. The model parameters fitted with $f= 
 0.3$ are quoted in Table 4; their variation with respect to $f=0$ is 
 within fit errors, implying that extra charm production does not 
 affect them significantly.
\begin{scriptsize}
\begin{table}
\caption[]{\footnotesize{Fit results for \ppb collisions at $\sqrt s= 900$ GeV with 
 $f_c =0.3$ compared with those without perturbative charm production.}} 
\begin{tabular}{lll}
\hline\noalign{\smallskip} 
   Parameter            & $f_c=0.3$          & $f_c =0$             \\ 
\noalign{\smallskip}
   Temperature(MeV)     & $163.8\pm10.9$     & $170.2\pm11.8$       \\	
   $VT^3$               & $46.0\pm12.1$      & $43.2\pm11.8$        \\ 
   $\gamma_s$           & $0.571\pm0.070$    & $0.578\pm0.063$      \\
   $\chi^2$/dof         & 3.09/2             & 1.79/2               \\
\noalign{\smallskip}\hline
\noalign{\smallskip}
\end{tabular}
\end{table}
\end{scriptsize}
 
\section{Fluctuations and correlations} 
 
 In the description of the model and the comparison of its 
 predictions with experimental data, we tacitly assumed that the 
 parameters $T$, $V$ and $\gamma_s$ do not fluctuate on an event by 
 event basis. If freeze-out occurs at a fixed hadronic density in 
 all events, as argued in Sect. 4, then it is a reasonable {\it 
 ansatz} that $T$ and $\gamma_s$ do not undergo any fluctuation since 
 the density mainly depends on those two variables. However, there 
 could still be volume fluctuations due to event by event 
 variations of the number and size of the fireballs from which the 
 primary hadrons emerge.\\ 
 We will now show that, as far as the average hadron multiplicities 
 are concerned, possible fluctuations of $V$ can be reabsorbed
 in a redefinition of volume provided that they are not too large. To 
 this end, let us define $\rho(V)$ as the probability density of 
 picking a volume between $V$ and $V+\dint V$ in a single event. The 
 primary average multiplicity of the $j^{th}$ hadron is then:
   
 \begin{equation}  
   \langle\!\langle n_j \rangle\!\rangle
      = \!\! \int \dint V \rho (V) \sum_{n=1}^{\infty} (\mp 1)^{n+1} 
      \, \gamma_s^{n s_j} z_{j(n)} \, \frac{Z(\QGz-n\qj)}{Z(\QGz)} \;.
 \end{equation} 
 If the volume $V$ fluctuates over a region where the dependence
 of chemical factors on it is mild (i.e. for large volumes, see Fig. 2), 
 they can be taken out of the integral in Eq.~(35) and evaluated at 
 the mean volume $\overline V$. In this case, the integrand depends 
 on the volume only through the functions $z_{j(n)}$ whose dependence 
 on $V$ is linear (see Eq.~(23)) and which can be re-expressed as

 \begin{equation}
 z_{j(n)}(V,T,\gamma_s) = V \, \xi_{j(n)}(T,\gamma_s)     \; .
 \end{equation}
 Then, from Eq.~(35),

 \begin{equation}  
   \langle\!\langle n_j \rangle\!\rangle \simeq \sum_{n=1}^{\infty} 
      (\mp 1)^{n+1} \, \gamma_s^{n s_j} \xi_{j(n)}(T,\gamma_s) \, 
      \frac{Z(\QGz-n\qj)}{Z(\QGz)} \int \dint V \rho (V) \, V  \; .
 \end{equation}
 The integral on the right-hand side is the mean volume $\overline V$. Thus:

 \begin{equation}  
    \langle\!\langle n_j \rangle\!\rangle \simeq \sum_{n=1}^{\infty} 
   (\mp 1)^{n+1} \, \gamma_s^{n s_j} z_{j(n)}(\overline V,T,\gamma_s) 
   \, \frac{Z(\QGz-n\qj)}{Z(\QGz)} \; .
 \end{equation} 
 It turns out that the relative hadron abundances do not depend on 
 the volume fluctuations since the mean volume $\overline V$ appearing in the 
 above equation (replacing the volume $V$ in Eq.~(23)) is the same for 
 all species: all results obtained in Sect. 4 are unaffected.
 On the other hand, if the volume fluctuates over a region where the 
 dependence of chemical factors on it is stronger (i.e. in the region
 of small volumes, see Fig. 2) one can show that the leading term of 
 average multiplicities is still given by the Eq.~(38) and that further 
 corrections are of the order of $D^2/\overline V^2$, where $D$ is the 
 dispersion of the distribution $\rho(V)$ (see Appendix E). Therefore, 
 if $D \ll \overline V$, as it is reasonably to be expected, also in this
 case the calculation of average multiplicities by using a single
 mean global volume would be a very good approximation.\\ 
 We emphasized in the Introduction that the average hadron 
 multiplicities are a very useful tool to study hadronization because 
 of their independence from collective dynamical effects. More 
 generally, since the number of particles is a Lorentz-invariant 
 quantity, this property is shared by the entire multiplicity 
 distribution of any hadron species. However, the shape of the 
 multiplicity distribution, unlike its mean value, is affected by 
 volume fluctuations since it is actually the superposition, weighted 
 with $\rho(V)$, of many multiplicity distributions, each of them 
 associated with a particular volume $V$, having different mean values 
 and moments. In a previous study \cite{giova} it has been shown that 
 the charged particle multiplicity distribution in \ee collisions at 
 $\sqrt s = 91.2$ GeV, calculated with a fixed volume, provides a fairly 
 good approximation of the experimental data, and that remaining 
 discrepancies between the prediction and the data can be explained
 by assuming a superposition of multiplicity distributions with 
 different volumes. This superposition effect (also called {\em 
 shoulder effect}) has been further investigated in ref. \cite{lupi}.\\ 
 Apart from the mean value, which is the first-order moment, the next 
 lowest order moments of multiplicity distribution are related to 
 global correlations between particle pairs. Let us first derive them 
 for a fixed volume $V$: let $n_{j,k}^i$ be the number of hadrons $j$ 
 in the $k^{th}$ phase space cell for the $i^{th}$ fireball; then, the 
 overall number of hadron $j$ is $\sum_{i,k} n_{j,k}^i$. According 
 to the partition function (18), the probability of picking a set of 
 occupation numbers $\{n_{j,k}^i\}$, i.e. a state of the system, is

 \begin{equation}
  P(\{n_{j,k}^i\}) = \frac{1}{Z} \, \exp (-\sum_{j,k,i} \beta_i 
   \cdot n_{j,k}^i p_k) \,  \delta_{\QG,\QGz}  \; .  
 \end{equation}
 The average number of pairs, whose first member belongs to species 
 $j$ and the second to species $l$, is then:

 \begin{equation}
   \langle\!\langle n_j n_l \rangle\!\rangle 
   = \sum_{\rm{states}} (\sum_{i,k} n_{j,k}^i)\, 
     (\sum_{i,k} n_{l,k}^i)\, P(\{n_{j,k}^i\})  
 \end{equation}
 if $j \neq l$, and

 \begin{equation}
    \langle\!\langle \frac{n_j (n_j-1)}{2} \rangle\!\rangle = 
       \frac{1}{2} \sum_{\rm{states}} (\sum_{i,k} n_{j,k}^i)\, 
       (\sum_{i,k} n_{j,k}^i-1)\, P(\{n_{j,k}^i\})    
 \end{equation} 
 if $j=l$. In both cases the average number of pairs can be obtained 
 from the partition function by multiplying all Boltzmann factors 
 $\exp \, (- \beta_i \cdot p_{k(j)})$ by fictitious fugacities 
 $\lambda_j$'s, one for each species, and taking the derivative with 
 respect to $\lambda_j$, $\lambda_l$ at ${\bf \lambda} = 1$: 

 \begin{equation}
    \langle\!\langle n_j n_l-\frac{1}{2}\delta_{jl}(n_j^2 + n_j) 
    \rangle\!\rangle  
    = (1-\frac{1}{2}\delta_{jl}) \, \frac {1}{Z} \frac {\partial^2 Z}
   {\partial \lambda_j \partial \lambda_l} \Big|_{{\bf \lambda} = 1}.  
 \end{equation} 
 Since the partition function is Lorentz-invariant and so are 
 the parameters $\lambda_j$, the average number of pairs does not 
 depend, as expected, on the collective fireball dynamics.\\ 
 In general, one can show that the average number of $n$-tuples of $K$ 
 particle species, with $n_1$ particles of species $j_1$, $n_2$ 
 particles of species $j_2$, $\ldots$, $n_K$ particles of species 
 $j_K$, is 
 
 \begin{equation}
   \frac{1}{n_1!\ldots n_K!}\frac{1}{Z}\frac{\partial^n Z}
  {\partial \lambda_1^{n_1} \ldots \partial \lambda_K^{n_K}} \; .
 \end{equation}    
 This expression proves that $Z (\lambda_1,\ldots,\lambda_K)$ is 
 proportional to the generating function of the multi-species 
 multiplicity distributions.\\ 
 The two-particle global correlation can be defined as the ratio 
 between the actual average number of pairs and the one that would have
 been obtained if their production was independent. Thus, if $j \neq l$: 

 \begin{equation}
   \rho_{jl} = \frac {\langle\!\langle n_j n_l \rangle\!\rangle}
    {\langle\!\langle n_j  \rangle\!\rangle 
     \langle\!\langle n_l  \rangle\!\rangle}  \; .
 \end{equation}
 As far as identical particles are concerned, if they were 
 independently produced their multiplicity distribution would be 
 Poissonian and therefore the average number of pairs would be 
 $\langle\!\langle n_j \rangle\!\rangle^2/2$, so: 
 
\begin{equation}
   \rho_{jj} = \frac{\langle\!\langle n_j^2-n_j \rangle\!\rangle}
                    {\langle\!\langle n_j \rangle\!\rangle^2} \; .
 \end{equation}
 The calculation of the average number of pairs according to Eq.~(42) and
 the partition function (20) yields:

 \begin{eqnarray}
  && \langle\!\langle n_j n_l-\frac{1}{2}\delta_{jl}(n_j^2 + n_j)
           \rangle\!\rangle = \nonumber \\
   &&  = (1-\frac{1}{2} \delta_{jl})  
    \sum_{m=1}^{\infty} \sum_{n=1}^{\infty} 
     (\mp 1)^{n+m} \, \gamma_s^{n s_j + m s_l} (z_{j(n)} z_{l(m)} 
       \mp \delta_{jl} z_{j(n+m)}) \frac{Z(\QGz-n\qj-m\ql)}{Z(\QGz)} \; ,
 \end{eqnarray}    
 where the upper sign is for fermions and the lower for bosons. 
 Whereas the term $z_{j(n)} z_{l(m)}$ is present for all particles, 
 the term $\delta_{jl} z_{j(n+m)}$ is non-zero only for identical 
 particles; it is a further contribution to correlated particle 
 production due to quantum statistics, the so called Bose-Einstein 
 correlations and Fermi-Dirac anticorrelations. If $j \!\ne\! l$ it 
 turns out that, comparing Eq.~(46) with Eq.~(22), $\langle\!\langle n_j 
 n_l \rangle\!\rangle  \ne \langle\!\langle n_l \rangle\!\rangle 
 \langle\!\langle n_l \rangle\!\rangle $ (and there is indeed 
 correlated production), unless $Z(\QGz-n\qj-m\ql) = Z(\QGz-
 n\qj)Z(\QGz-m\ql)/Z(\QGz)$; this is the case if the function $Z(\QG)$ 
 is an exponential of $\QG$, which occurs only in the grand-canonical 
 regime (see also Appendix B). Thus, in the canonical thermodynamical 
 approach, correlated production of particles belonging to different 
 species is definitely an effect of conservation laws in a finite 
 system. As long as different species are concerned, owing to the 
 temperature values found in the present analysis the contribution to 
 the average number of pairs from terms other than $n=1$ and $m=1$ in 
 series (46) is negligible for all hadrons but pions. Therefore:
  
 \begin{equation}
   \langle\!\langle n_j n_l \rangle\!\rangle  \simeq \gamma_s^{s_j} 
   \gamma_s^{s_l} z_j z_l \, \frac{Z(\QGz-\qj-\ql)}{Z(\QGz)} \; , 
 \end{equation}    
 which corresponds to the Boltzmann limit, as discussed in Sect. 3. On 
 the other hand, for all identical particles but pions we have:

 \begin{equation}
   \langle\!\langle \frac{n_j (n_j -1)}{2} \rangle\!\rangle 
   \simeq \frac{1}{2} \gamma_s^{2s_j} 
   (z_j^2 \mp z_{j(2)})\, \frac{Z(\QGz-2\qj)}{Z(\QGz)} \; .
 \end{equation}  
 In principle, the term $z_{j(2)}$ stemming from quantum statistics 
 may not be negligible compared to $z_j^2$ even for high mass hadrons, 
 since the ratio 

 \begin{equation}
  \frac{z_{j(2)}}{z_j^2} = \frac{1}{2 z_j} \frac{{\rm K}_2(2m_j/T)}
   {{\rm K}_2(m_j/T)}
 \end{equation}
 may be of the order of 1 if, due to a very small volume, $z_j \ll 1$ 
 is able to compensate the small ratio of McDonald functions. 
 In the present analysis the largest value for the ratio (49) for heavy 
 hadrons (i.e. excluding pions) which occurs is 0.096 for 
 K$^{+}$K$^{+}$ production in pp collisions at $\sqrt s = 19.4 $ GeV.\\ 
 The global correlation between heavy hadron pairs turns out to be, 
 using Eqs.~(44)-(48) and (29), 
 
 \begin{equation}
   \rho_{jl} \simeq (1 \mp \, \delta_{jl} \frac{z_{j(2)}}{z_j^2}) 
   \frac{Z(\QGz-\qj-\ql)Z(\QGz)}{Z(\QGz-\qj)Z(\QGz-\ql)} \; . 
 \end{equation} 
\begin{scriptsize}
\begin{table*}
\caption[]{\footnotesize{Particle-particle correlations in pp collisions 
  at $\sqrt s = 26$ GeV \cite{corr} obtained by using the total inelastic cross
  section of 32.80 mb quoted in ref. \cite{asai}. The errors within
  brackets next to the theoretical predictions are due to finite 
  Monte-Carlo statistics. The experimental values of the correlations 
  $\rho$ have been estimated by dividing the average numbers of pairs 
  by the average multiplicities quoted in ref. \cite{asai}, measured in the same 
  experiment. Since the correlation between these two measurements is 
  unknown, the relative experimental error on $\rho$ has been assumed to be 
  the same as on $\langle\!\langle n_1 n_2 \rangle\!\rangle$. The 
  small effect of quantum statistics on the correlations of identical 
  particles at the primary hadron level has been neglected, as explained 
  in the text.}} 
\begin{tabular}{llllll}
\hline\noalign{\smallskip} 
   Particles            &$\langle n_1 n_2 \rangle$ measured&
                         $\langle n_1 n_2 \rangle$ calculated&
                         $\rho$ measured&$\rho$ calculated  &
                         $\rho$ for primaries \\ 
\noalign{\smallskip}
   K$^0_s$K$^0_s$       &$0.0530\pm0.0055$   & $0.0489 (\pm0.0021)$ &$1.57\pm0.16$  & $1.49  (\pm0.064)$&  1.44$(\pm0.11)$\\
   K$^0_s$$\Lambda$     &$0.0451\pm0.0052$   & $0.0472 (\pm0.0018)$ &$1.45\pm0.17$  & $1.25  (\pm0.048)$&  1.506           \\  
   K$^0_s$$\bar\Lambda$ &$0.0079\pm0.0021$   & $0.0041 (\pm0.0006)$ &$2.34\pm0.62$  & $1.34  (\pm0.18)$ &  1.296           \\ 
   $\Lambda\bar\Lambda$ &$0.0030\pm0.0009$   & $0.0036 (\pm0.0004)$ &$1.92\pm0.58$  & $2.05  (\pm0.25)$ &  2.820           \\ 
\noalign{\smallskip}\hline
\noalign{\smallskip}
\end{tabular}
\end{table*}
\end{scriptsize}
 All calculations performed in this Section refer to primary 
 hadrons, which are not observable in actual experiments. Since 
 measured correlations may be affected by the decay chain 
 process, the given formulae are not directly comparable with 
 experimental data. Therefore, a complete reconstruction of the 
 production process including both the formation and decay of the
 primary hadrons is necessary in order to test the predictive 
 power of the model in this regard. This can be done by a Monte-Carlo 
 procedure: by using the model parameters fitted at each centre of 
 mass energy point as described in Sect. 4, a set of numbers 
 $\{n_{j,k}^i\}$ is generated according to the probability (39), and 
 subsequently their decays are performed according to the known decay 
 modes and branching ratios as quoted in the Particle Data Book 
 \cite{pdg}.\\ 
 A further problem in the comparison with data is the possibility of 
 volume fluctuations. If the volume fluctuates from event to event,
 the average number of heavy hadron pairs should be re-expressed as 

 \begin{equation}
    \langle\!\langle n_j n_l-\frac{1}{2}\delta_{jl}(n_j^2 + n_j)
     \rangle\!\rangle \simeq \int \dint V \rho(V) \, \gamma_s^{s_j + s_l} 
      (z_j z_l \mp \delta_{jl} 
      \, z_{j(2)}) \, \frac{Z(\QGz-\qj-\ql)}{Z(\QGz)} .
 \end{equation}   
 According to what has been stated in the beginning of this section, 
 if we take the chemical factors out of the integral and write $z = 
 V \xi (T,\gamma_s)$, $z_{(2)}= V \xi_{(2)}(T,\gamma_s)$, we are left 
 in Eq.~(49) with both a mean volume (multiplying $\delta_{jl}$) and a 
 mean squared volume which is equal to the squared mean volume only if 
 the dispersion $D$ of the distribution $\rho(V)$ vanishes. Taking 
 into account volume fluctuations, the correlation $\rho_{jl}$ reads 
 (taking the first term of the series in Eq.~(38)):   

 \begin{equation}
   \rho_{jl} = \left(\frac{\overline{V^2}}{{\overline V}^2} \mp \, \delta_{jl} 
    \frac{z_{j(2)}}{z_j^2}\right) \frac{Z(\QGz-\qj-\ql)Z(\QGz)}{Z(\QGz-\qj)Z(\QGz-\ql)} 
   = \left(1+ \frac{D^2}{{\overline V}^2} \mp \, \delta_{jl} \frac{z_{j(2)}}{z_j^2}\right) 
    \frac{Z(\QGz-\qj-\ql)Z(\QGz)}{Z(\QGz-\qj)Z(\QGz-\ql)}  \; .
 \end{equation}   
 The correlation between different particle species then increases by 
 a factor $(1+ D^2/{\overline V}^2)$ with respect to the non-fluctuation
 case even neglecting the dependence of chemical factors on the volume; 
 however, if $D \ll \bar V$ this a small effect.\\ 
 We compared Monte-Carlo simulated correlations with a set of 
 particle-particle correlations measured in pp collisions at $\sqrt s 
 = 26.0$ GeV \cite{corr}; the results are shown in Table 5. The 
 correlations have been predicted by using the model parameters $T$, 
 $V$, and $\gamma_s$ fitted at that centre of mass energy quoted in 
 Table 1. We also quote the correlations at the primary hadron level 
 which were calculated with Eqs.~(47) and (48), taking into account 
 that K$^0_s$ is a mixed particle-antiparticle state, and, for the 
 K$^0_s$K$^0_s$ correlation, with the same Monte-Carlo technique used 
 for the final particles. The effect of Bose-Einstein correlations in 
 the global correlated production of K$^0_s$K$^0_s$, estimated to be 
 $< 1.7 \%$ according to formulae (48), (49) and taking mixing into 
 account, has been neglected. The comparison between primary and final 
 correlations indicates that, in general, they are slightly diluted by 
 the decay chain process.\\
 Generally, the agreement between the predictions and the data is 
 good. In trusting this approach one is led to the conclusion that 
 volume fluctuations are small enough to be hidden in the experimental 
 errors. This fact should be confirmed by a more detailed study of 
 charged particle multiplicity distributions.  
 
\section{Conclusions} 	

 A detailed analysis of hadron abundances in pp and \ppb collisions 
 over a large range of centre of mass energies (from 
 20 to 900 GeV) has demonstrated a stunning ability of the 
 thermodynamic model to reproduce accurately all available 
 experimental data on hadron production in high energy collisions 
 between elementary hadrons. Key elements for the success of this 
 approach are the use of the canonical formalism of statistical 
 mechanics, ensuring the exact implementation of quantum number 
 conservation, and the introduction of a supplementary parameter 
 $\gamma_s$ to account for incomplete saturation of strange 
 particle phase space.\\
 The remarkable agreement of the data with such a purely statistical 
 approach, which uses only three free parameters, has important 
 implications. Firstly, it indicates that hadron production in elementary 
 high energy collisions is dominated by phase space rather than by 
 microscopic dynamics; during hadronization of the prehadronic matter 
 formed in the collision, the hadronic phase space is filled according 
 to the law of maximal entropy, with minimal additional (i.e. 
 dynamical) information. The only dynamics visible in the final state
 is the collective motion of the hadron gas fireballs which reflects 
 the underlying hard parton kinematics. Secondly, the observed universality 
 of the freeze-out temperature independent of the collision energy and 
 collision system suggests that 
 hadronization cannot occur before the parameters of prehadronic matter, 
 like energy density or pressure, have dropped below critical values 
 corresponding to a temperature of around 170 MeV in an (partially) 
 equilibrated hadron gas.
 Hadronization at larger energy densities or pressures is inhibited 
 by the ``absence" of a hadronic phase space: according to lattice QCD 
 calculations, the most likely (i.e. maximum entropy) state at higher 
 energy densities is a colour deconfined quark-gluon plasma in which 
 hadrons do not exist as stable degrees of freedom. Therefore,
 this analysis indicates that the value of critical transition 
 temperature is $T_{\rm crit} \simeq 170$ MeV. This agrees with the 
 limiting (``Hagedorn") temperature \cite{hag,hag2} for
 an equilibrated hadron gas and with lattice QCD results \cite{qm96}.\\ 
 The phase-space dominance in the hadronization process can be 
 understood by the non-perturbative nature of the strong interaction 
 forces in this energy density domain: within each fireball, many different 
 processes and channels contribute to the formation of soft hadrons, 
 resulting locally in equal transition probabilities for all hadronic
 states in phase space. The value of temperature reflects the hadronic
 energy density or pressure at its critical value where hadron
 production occurs while the only other parameter entering the 
 observed hadron spectra is the collective motion relative to 
 the observer.\\
 The only deviation from this picture of complete phase-space 
 dominance in hadronization resides in the incomplete saturation of 
 strange particle phase space, i.e. $\gamma_s \simeq 0.5$: the final 
 hadronic state seems to ``remember" that there were no strange quarks 
 in the initial state, and, in spite of their non-perturbative nature and
 the many possible dynamical channels, strong interactions
 in the pre-hadronic stage do not manage to wipe out completely the
 asymmetry between strange quark and light quark abundances.
 Strangeness suppression, as well as the survival of perturbatively 
 created $\cc$ and $\bb$ pairs, are thus the only trace to strong 
 interaction dynamics before hadronization. The systematics of the 
 observations suggest that these non-equilibrium effects are mainly 
 related to quark mass thresholds. Similar analyses of hadron 
 abundances in nuclear collisions suggest that the strangeness 
 suppression disappears in larger collision systems with larger
 lifetimes prior to hadron freeze-out \cite{hein3}. Note that also 
 here, in hadronic collisions, the slight increase of $\gamma_s$ 
 with centre of mass energy in hadronic collisions is connected with a 
 systematic increase of the fitted fireball volumes at freeze-out 
 (see Table 1); this goes in the same direction. The increase of 
 freeze-out volume with rising centre of
 mass energy may imply a corresponding increase of the initial 
 (prehadronic) energy density since the final energy density of the 
 hadronic state is limited by the observed constant temperature.\\
 It has been shown that not only the average single hadron 
 abundances, but also the particle-particle correlations measured in 
 pp collisions agree well with the thermal predictions. It should be 
 stressed that the requirement of exact quantum number 
 conservation yields major effects on both the average hadron 
 multiplicities and the correlations, and that for elementary high 
 energy collisions a thermal description in the grand-canonical 
 framework would not have worked so well.
 As noted by Hagedorn many years ago \cite{hag2,hag3} and extensively
 discussed in Sect. 2, small fireballs volume result in the 
 suppression of strange relative to non-strange hadrons even for $\gamma_s=1$
 (i.e. without strange phase space suppression), when one compares 
 the canonical with the grand-canonical approach, due to the need
 to create strange particles always in pairs.  
 A strangeness suppression $\gamma_s \simeq 0.5$ in the canonical 
 approach, as extracted here from the pp and \ppb data, thus 
 corresponds to a seemingly much stronger suppression of $\gamma_s 
 \simeq 0.2$ within a grand-canonical approach \cite{hein3}. Vice-versa, 
 it should be emphasized that even a constant value of $\gamma_s$ 
 may imply a strong relative enhancement of strange particles going from
 the small volumes of elementary hadron collisions to possible large 
 volumes in nuclear collisions. The frequently discussed ``strangeness
 enhancement" in nuclear collisions thus really consists of two
 components: (i) the removal of the suppression (at constant $\gamma_s$) 
 arising from the need to conserve exactly strangeness in a small 
 collision volume, and on top of that (ii) additionally a possibly 
 larger value of $\gamma_s$ \cite{heinz2,hein3}. Both of these effects 
 are dynamically non-trivial. 

\section*{Acknowledgement}
 Stimulating discussions with A. Giovannini, R. Hagedorn and H. Satz are 
 gratefully acknowledged. One of the authors (U. H.) would like to thank 
 the CERN theory group for warm ospitality during his sabbatical. 
 Thanks to J. A. Baldry for the careful revision of the manuscript. 

\newpage

\section{Appendix} 
\appendix 	
\section{Proof of Equation (20)} 
 
 We want to prove that the global partition function (18) can be expressed by Eq.~(20) 
 if the temperatures and the strangeness suppression factors $\gamma_s$ of the 
 various fireballs are constant. Let $n_{j,k}^i$ be the number of the $j^{th}$ 
 hadron species in the $k^{th}$ phase space cell of the $i^{th}$ fireball. Then:

 \begin{eqnarray}
 \!\!\!&& \sum_i \QGi = \sum_{i,j,k} n_{j,k}^i \qj   \nonumber \\
 \!\!\!&& P_i          = \sum_{j,k} n_{j,k}^i p_k   \; . 
 \end{eqnarray}
 By using Eq.~(2) and putting Eq.~(53) into Eq.~(18) the following expression of
 global partition function is obtained: 

 \begin{equation}
     Z(\QGz) = \frac{1}{(2\pi)^5} \int \, \dint^5 \phi \,\, \E^{\,\I\, \QGz \cdot \phi} 
       \prod_{i=1}^N \sum_{\rm{states}_i} \, 
     \exp \, [- \sum_{j,k} \beta_i \cdot n_{j,k}^i p_k - \,\I\, n_{j,k}^i \qj \cdot \phi ]  \; .
 \end{equation} 
 After summing over states and inserting the strangeness suppression factor
 $\gamma_s$, the Eq.~(54) becomes:

 \begin{equation}
    Z(\QGz) = \frac{1}{(2\pi)^5} \int \, \dint^5 \phi \,\, 
    \E^{\,\I\, \QGz \cdot \phi} \prod_{i=1}^N \exp \, [\sum_{j} \sum_{k} \, \log \, 
   (1 \pm \gamma_s^{s_j}\E^{-\beta_i \cdot p_{k} -\I \qj \cdot \phi})^{\pm1}]  \; ,
 \end{equation}
 where the upper sign is for fermions and the lower for bosons.\\  
 Once the transformation (6) has been applied in Eq.~(55), one is left with phase 
 space integrals that may be performed in the rest frame of each fireball, in the 
 very same way as in Eq.~(7):
 
 \begin{equation}
   Z(\QGz)= \frac{1}{(2\pi)^5} \int \, \dint^5 \phi \,\, \E^{\,\I\, \QGz \cdot \phi} 
   \prod_{i=1}^N \exp \, [V_i \sum_j F_j(T_i,\gamma_{si},\PG)] \; . 
 \end{equation}   
 If $T_1=\ldots=T_N \equiv T$ and $\gamma_{s1}=\ldots=\gamma_{sN} \equiv \gamma_s$, 
 then:

 \begin{equation}
  Z(\QGz)= \frac{1}{(2\pi)^5} \int \, \dint^5 \phi \,\, \E^{\,\I\, \QGz \cdot \phi} 
   \exp \, [(\sum_i V_i) \sum_j F_j(T,\gamma_{s},\PG)] \;.         
 \end{equation}
 which is precisely the Eq.~(20).

\section{Approximation of the function $Z(\QG)$ for large systems} 	

 We look for an approximated expression of the function $Z(\QG)$ for large values
 of particle multiplicity, namely for large values of volume $V$. In the 
 following calculations heavy flavoured particles, whose multiplicities are 
 orders of magnitude below light flavoured ones, at temperatures $T = {\cal O}
 (100)$ MeV, are completely neglected. This means that we are dealing with a
 function $Z(\QG)$ as in Eq.~(32) and that vectors $\QG$ and $\qj$ are 
 henceforth meant to be three-dimensional with components electric charge, baryon 
 number and strangeness respectively.\\
 Let us define:

 \begin{equation}
  f(\PG) \equiv 
    \sum_j \frac{2J_j+1}{(2\pi)^3} \int \dint^3 p \, \log \, (1 \pm \gamma_s^{s_j} 
      \E^{-\sqrt{p^2+m_j^2}/T -\I \qj \cdot \phi})^{\pm 1} \; , 
 \end{equation}
 where the upper sign is for fermions and the lower is for bosons. By using this
 definition, the function $\zeta(\QG)$ in Eq.~(32) can be written: 

 \begin{equation}
   \zeta(\QG) = \frac{1}{(2\pi)^3} \int \, \dint^3 \phi \,\, \E^{\,\I\, \QG \cdot \phi}
     \exp \, [V f(\PG)]  \; . 
 \end{equation}
 The logarithm in the function $f(\PG)$ in Eq.~(58) can be expanded in a series: 

 \begin{equation}
     \log \, (1 \pm \gamma_s^{s_j} \E^{\sqrt{p^2+m_j^2}/T} 
       \E^{-\I \qj \cdot \phi})^{\pm 1}  \nonumber \\ 
      = \sum_{n=1}^\infty \frac{(\mp 1)^{n+1}}{n}\,\gamma_s^{n s_j} 
       \E^{-n \sqrt{p^2+m_j^2}/T} \E^{-n \,\I\, \qj \cdot \phi} \; . 
 \end{equation}
 The sum labelled by index $j$ in the function $f(\PG)$ obviously runs over all
 particles and anti-particles species. However, in order to develop calculations, 
 it is advantegeous to group particles and corresponding anti-particles terms together 
 in the series (60). Doing that, the Eq.~(58) becomes:

 \begin{equation}
     f(\PG) = \!\!\!\!\! \sum_{j \pa} \!\!\! \frac{2J_j+1}{(2\pi)^3} 
        \int \dint^3 p \sum_{n=1}^\infty \frac{(\mp 1)^{n+1}}{n} \,
     2 \, \gamma_s^{n s_j} \E^{-n \sqrt{p^2+m_j^2}/T} \cos\,(-n \qj \cdot \PG) \; .  
 \end{equation}
 Since the integrand function in Eq.~(59) is periodical, the integration can be 
 performed in the interval $[-\pi,\pi]$ instead of $[0,2\pi]$. The reason of this 
 shift in the integration interval is that
 a considerable property of the function $f(\PG)$ is the presence of a maximum 
 at $\PG=0$, and, consequently, a very peaked maximum in the same point for the 
 function $\exp\,[Vf(\PG)]$ for large values of $V$. In this case, the 
 saddle-point approximation can be used in calculating the integral (59). 
 Therefore:

 \begin{eqnarray}
   f(\PG) &\simeq& \!\!\!\!\! \sum_{j \pa} \!\!\! \frac{2J_j+1}{(2\pi)^3} 
      \int \dint^3 p \sum_{n=1}^\infty \frac{(\mp 1)^{n+1}}{n} \, 2 \, \gamma_s^{n s_j} 
       \E^{-n \sqrt{p^2+m_j^2}/T} [1- (n \qj \cdot \PG)^2/2] \nonumber \\
  &=& f(0) - (\qj \cdot \PG)^2  
      \!\!\!\!\! \sum_{j \pa} \!\!\! \frac{2J_j+1}{(2\pi)^3} \int \dint^3 p \, 
     \frac{\gamma_s^{s_j} \E^{-\sqrt{p^2+m_j^2}/T}}
     {(1\pm\gamma_s^{s_j} \E^{-\sqrt{p^2+m_j^2}/T})^2} \; . 
 \end{eqnarray} 
 Let us define now a $3\times 3$ real symmetric matrix $\sf{A}$ whose elements are:

 \begin{equation}
 {\sf{A}}_{k,l} = \!\!\!\!\!\!\! \sum_{j \pa} \!\!\! \frac{V \, 
    (2J_j+1)}{(2\pi)^3} \int \dint^3 p \,\, \frac{\gamma_s^{s_j} \E^{-\sqrt{p^2+m_j^2}/T}}
    {(1\pm\gamma_s^{s_j}\E^{-\sqrt{p^2+m_j^2}/T})^2} \,\, q_{j,l} q_{j,k} \; .
 \end{equation}  
 By using this definition, the Eq.~(62) reads:

 \begin{equation}
   f(\PG) \simeq f(0) -  \PG \cdot \frac{{\sf{A}}}{V} \PG  \; .
 \end{equation}   
 Thus:

 \begin{equation}
   \zeta(\QG) \simeq \frac{1}{(2\pi)^3} \exp \, [V f(0)] \int \, \dint^3 \phi 
   \,\, \E^{\,\I\, \QG \cdot \phi} \exp \, [- \PG \cdot {\sf{A}} \PG] \; .
 \end{equation}   
 If $V$ is large enough, the integration can be extended from $[-\pi, \pi]$ to 
 $[-\infty,\infty]$ without affecting significantly the final result. Hence:
 
 \begin{equation}
   \zeta(\QG) \simeq \frac{1}{(2\pi)^3} \exp [V f(0)] 
   \sqrt{\frac{\pi^3}{\det{\sf{A}}}} \exp [-\frac{1}{4} \QG {\sf{A}}^{-1} \QG ] \; .  
 \end{equation}
 Now we are able to write an approximate expression of chemical factors in the Eq.~
 (22):
 
 \begin{equation}
   \frac{Z(\QG-n\qj)}{Z(\QG)} = 
    \frac{\exp [-\frac{1}{4}(\QG-n\qj){\sf{A}}^{-1}(\QG-n\qj)]}
   {\exp [-\frac{1}{4}\QG{\sf{A}}^{-1}\QG]} = \exp[\frac{n}{2}  
    \QG {\sf{A}}^{-1} \qj] \exp [ - \frac{n^2}{4} \qj {\sf{A}}^{-1} \qj] \; .
 \end{equation}  
 By using this approximation, the average multiplicity of primary hadrons (29) 
 in the Boltzmann limit can now be written as:

 \begin{equation}
   \langle\!\langle n_j \rangle\!\rangle = (2J_j+1) \, \frac{V}{(2\pi)^3} \,
      \gamma_s^{s_j} \int \dint^3 p \,\, \E^{-\sqrt{p^2+m_j^2}/T} \,\, 
     \E^{\QG {\sf{A}}^{-1} \qj/2} \,\, \E^{-\qj {\sf{A}}^{-1} \qj/4} \; .
 \end{equation}
 To summarize, in the large volume limit, chemical factors reduce to a product 
 of two factors: the first corresponds to a traditional chemical potential 
 whereas the second does not have a corresponding grand-canonical quantity; 
 its presence is ultimately due to internal (i.e. quantum numbers) conservation
 laws in a finite system. Since:

 \begin{equation}
  \lim_{V \rightarrow \infty} {\sf{A}}^{-1} = 0  
 \end{equation}
 the additional suppression factor $\exp[-\qj {\sf{A}}^{-1} \qj/4]$ is 
 negligible in the proper thermodynamic limit provided that vectors $\qj$ 
 are finite: the grand-canonical formalism is recovered.

\section{Heavy flavoured hadrons production} 	

As shown in Sect. 4, the average multiplicity of primary charmed hadrons in 
events in which one ${\rm{c\bar c}}$ pair is created owing to a hard QCD 
process must be calculated with the usual Eq.~(28) in which the partition function 
$Z$ is (see Eq.~(33)):

\begin{equation}
 Z = Z_1(\QGz)-Z_2(\QGz,0) \; .
\end{equation}
The function $Z_1$ can be written in the very same fashion as in Eq.~(31):

 \begin{equation}
   Z_1(\QGz) \simeq \frac{1}{(2\pi)^5} \int \, \dint^5 \phi \,\, \E^{\,\I\, \QGz \cdot \phi} 
       \exp \, [ \sum_j z_j \gamma_s^{s_j} \E^{-\I \qj \cdot \phi} 
      + \sum_{j=1}^{3} \frac{V}{(2\pi)^3} \int \dint^3 p \, 
      \log \, (1 - \E^{-\sqrt{p^2+m_j^2}/T -\I \qj \cdot \phi})^{-1}] \; ,
 \end{equation} 
while the function $Z_2$ can be worked out according to the same procedure depicted 
for the function $Z_2$ in Eqs.~(26), (27) for the leading baryon effect:

 \begin{eqnarray}
   Z_2(\QGz,K) &=& \frac{1}{(2\pi)^6} \int  \dint^5 \phi \,\, \E^{\,\I\, \QGz \cdot \phi}
     \int \dint \psi 
      \,\, \E^{\,\I\, K \psi}  \nonumber \\
  &\times& \exp \, [ \sum_{j=1} z_j \gamma_s^{s_j} \E^{-\I \qj \cdot \phi -\I |C_j| \psi}
         + \sum_{j=1}^{3} \frac{V}{(2\pi)^3} \int \dint^3 p \, 
      \log \, (1 - \E^{-\sqrt{p^2+m_j^2}/T -\I \qj \cdot \phi})^{-1}] \; , \nonumber \\
  &&
\end{eqnarray}
where the second sum in the exponentials in both Eqs.~(71) and (72) runs over the 
three pion states and $|C_j|$ in Eq.~(72) is the absolute value of $j^{th}$ hadron's 
charm.\\ 
Henceforth, we denote by $\QGz$ and $\qi$, $\qj$, $\qk$ three-dimensional vectors 
having as components electric charge, baryon number and strangeness, while charm and
beauty will be explicitely written down. By using this notation, the average multiplicity 
of a charmed hadron with $C_j = 1$ turns out to be (the Boltzmann limit holds, 
cf. Eq.~(29)):

\begin{equation}
      \langle\!\langle n_j \rangle\!\rangle 
          =  z_j \frac{Z_1(\QGz-\qj,-1,0) - Z_2(\QGz-\qj,-1,0,-1)}
                      {Z_1(\QGz,0,0) - Z_2(\QGz,0,0,0)} \; . 
\end{equation}
Since the $z$ functions of heavy flavoured hadrons are $\ll 1$, as shown in 
Sect. 4, a power expansion in the $z_j$'s of all charmed and anti-charmed hadrons 
can be performed from $z_j = 0$ in the integrands of Eqs.~(71) and (72), that is: 

\begin{equation}
    \exp [\sum_j \gamma_s^{s_j} z_j \E^{-\I \qj \cdot \phi}] 
          \simeq 1 + \sum_j \gamma_s^{s_j} z_j \E^{-\I \qj \cdot \phi} 
      + \frac{1}{2} \sum_{i,j} \gamma_s^{s_i} \gamma_s^{s_j} z_i z_j 
       \E^{-\I (\qj+\qi) \cdot \phi}   
\end{equation}
for Eq.~(71) and

\begin{equation}
   \exp \, [\sum_j \gamma_s^{s_j} z_j \E^{-\I \qj \cdot \phi -\I |C_j| \psi}] 
 \simeq  1 + \sum_j z_j \E^{-\I \qj \cdot \phi - \,\I\, |C_j| \psi}
   + \frac{1}{2} \sum_{i,j} \gamma_s^{s_i} \gamma_s^{s_j} z_i z_j 
   \E^{-\I (\qj+\qi) \cdot \phi -2 \I \, |C_j| \psi} 
\end{equation}
for Eq.~(72). Furthermore, the $z$ functions of the bottomed hadrons can be 
neglected as they are $\ll 1$ as well and beauty in Eq.~(73) is always set to zero.\\
Those expansions permit carrying out integrations in the variables $\psi$, 
$\phi_4$ and $\phi_5$ in Eqs.~(71) and (72). Thus:

\begin{equation}
  Z_1(\QGz-\qj,-1,0) \simeq \sum_i \gamma_s^{s_i} z_i \zeta(\QGz-\qj-\qi)  \; , 
\end{equation}
where the sum runs over the {\em anti-charmed hadrons} as the integration in $\phi_4$ 
of terms associated to charmed hadrons yields zero. The $\zeta$ function on the 
right-hand side is the same as in Eq.~(32). Moreover:

\begin{equation}
  Z_1(\QGz,0,0) \simeq \zeta(\QGz) + \sum_{i,k} \gamma_s^{s_i} \gamma_s^{s_k}
  z_i z_k \zeta(\QGz-\qi-\qk) \; ,
\end{equation}
where the index $i$ runs over all charmed hadrons and index $k$ over all anti-charmed
hadrons.\\
Owing to the presence of the absolute value of charm in the exponential 
$\exp[\,\I\, |C_j| \psi]$ ($|C_j| = 1$) in its integrand function, the 
function $Z_2(\QGz,C,B,K)$ 
vanishes if $K \le 0$ and yields $K^{th}$-order terms of the power expansion in 
$z_j$ if $K \ge 0$ (see Eq.~(72)). Therefore: 
 
\begin{equation}
  Z_2(\QGz,0,0,0) = \zeta(\QGz)  
\end{equation}
and

\begin{equation}
  Z_2(\QGz,-1,0,-1) = 0  \; .
\end{equation}
Finally, inserting Eqs.~(76), (77), (78) and (79) in Eq.~(73) one gets: 

\begin{equation}
  \langle\!\langle n_j \rangle\!\rangle 
  = \gamma_s^{s_j} z_j \,\, \frac{\sum_i \gamma_s^{s_i }z_i \zeta(\QGz-\qj-\qi)}
   {\sum_{i,k} \gamma_s^{s_i} \gamma_s^{s_k} z_i z_k \zeta(\QGz-\qi-\qk)} \; ,
\end{equation}
where the indices $j$, $k$ label charmed hadrons and $i$ labels anti-charmed hadrons. 
From previous equation it results that the overall number of primary charmed 
hadrons is 1, as it must be if c quark production from fragmentation is negligible.
The average multiplicity of anti-charmed hadrons is of course equal to charmed 
hadrons one. The same formula (80) holds for the average multiplicity of 
bottomed hadrons in events with a perturbatively generated ${\rm{b\bar b}}$ pair.
If leading baryon effect is taken into account, the formula (80) gets more 
complicated, but the procedure is essentially the same.\\
It is clear that a possible charm or beauty suppression parameter $\gamma_c$ 
or $\gamma_b$, introduced by analogy with strangeness suppression parameter 
$\gamma_s$, would not be revealed from the study of heavy flavoured hadron production
because a single factor multiplying all $z_j$ functions would cancel from the 
ratio in the right-hand side of Eq.~(80).

\section{On the definition of pressure} 	

In Sect. 4, we have dealt with pressure as a single well-defined 
quantity for the whole system of hadron gas fireballs. However, since the system 
has local collective flows, the definition of a single pressure is not a
trivial one. We will now show that the best (as well as the most natural)
definition is:

\begin{equation}
p = T \, \frac{\partial \log Z}{\partial V} \; ,
\end{equation}
where $Z$ is the global partition function (see Eqs.~(18)-(20)) and $V$ the global
volume defined in Sect. 2.\\
If the temperatures and $\gamma_s$ parameters of the fireballs are the same,
as we have assumed throughout, the global partition function depends on the 
single fireball volumes only through the sum $\sum_{i=1}^N V_i$, as shown in 
Eq.~(20). Therefore we can replace the derivative in Eq.~(81) with:

\begin{equation}
p = T \sum_{i=1}^N \frac{V_i}{V} \, \frac{\partial \log Z}{\partial V_i} \; .
\end{equation}
In order to develop this equation, we can use the expression (15) of the global 
partition function and write:

\begin{equation}
p = \frac{T}{V} \sum_{i=1}^N V_i \, \frac{\partial}{\partial V_i} \log \!\!\!\!
 \sum_{\QG_1^0,\ldots,\QG_N^0} \!\!\!\! \delta_{\zum_j \QGzj,\QGz} 
                    \prod_{j=1}^N Z_j(\QGzj) \; , 
\end{equation}
which is equal to:

\begin{equation}
 p = \frac{T}{V} \sum_{i=1}^N V_i \!\!\!\! \sum_{\QG_1^0,\ldots,\QG_N^0} \!\!\!\!
 w(\QG_1^0,\ldots,\QG_N^0) \frac{\partial}{\partial V_i} \log \prod_{j=1}^N 
 Z_j(\QGzj) 
\end{equation}
by using the weights defined in Eq.~(13). It is now possible to expand
the derivative in Eq.~(84): 

\begin{equation}
 \frac{\partial}{\partial V_i} \log \prod_{j=1}^N Z_j(\QGzj) =
 \sum_{j=1}^N \frac{\partial}{\partial V_i} \log Z_j(\QGzj)
 =\frac{\partial}{\partial V_i} \log Z_i(\QGzi) \; ;
\end{equation}
the last equality is due to the dependence of $Z_j(\QGzj)$ only on the volume 
$V_j$.\\
We can now write the pressure as:

\begin{equation}
p = T \!\!\!\! \sum_{\QG_1^0,\ldots,\QG_N^0} \!\!\!\! w(\QG_1^0,\ldots,\QG_N^0)
 \sum_{i=1}^N \frac{V_i}{V} \, \frac{\partial}{\partial V_i} \log Z_i(\QGzi) \; .
\end{equation}
The expression $T \partial \log Z_i(\QGzi)/\partial V_i$ in the 
above equation is the pressure $p_i(\QGzi)$ of the $i^{th}$ fireball, which 
is a well-defined one, as the fireball is a system at complete thermal and
mechanical equilibrium by definition. Therefore the global pressure turns 
out to be:

\begin{equation}
p = \!\!\!\! \sum_{\QG_1^0,\ldots,\QG_N^0} \!\!\!\! w(\QG_1^0,\ldots,\QG_N^0) 
 \sum_{i=1}^N \frac{V_i}{V} \, p_i(\QGzi) \; .
\end{equation}
The last equation now makes it clear that the definition (81) is the most 
natural definition of pressure for two reasons:
\begin{enumerate}
\item for a given event, in which a specific configuration of fireball quantum
numbers $\{\QG_1^0,\ldots,\QG_N^0\}$ is created, the global pressure 
is the average of the pressures of single fireballs, weighted by their 
extension through the factor $V_i/V$;
\item in general, the global pressure is the average over 
all possible configurations of fireball quantum numbers according to their 
probabilities of occurrence $w(\QG_1^0,\ldots,\QG_N^0)$.
\end{enumerate} 

\section{Average multiplicities and volume fluctuations} 	

We want to calculate the leading correction to the formula (38) for particle
average multiplicities in presence of volume fluctuations affecting chemical 
factors $Z(\QGz-n\qj)/Z(\QGz)$. According to Eqs.~(35), (36):

\begin{eqnarray}  
  \!\!\! && \langle\!\langle n_j \rangle\!\rangle
      = \!\! \int \dint V \rho (V) \sum_{n=1}^{\infty} (\mp 1)^{n+1} 
      \, \gamma_s^{n s_j} V \xi_{j(n)} f_{j(n)} \; , \nonumber \\ 
  \!\!\! && 
\end{eqnarray}    
where $f_{j(n)} \equiv Z(\QGz-n\qj)/Z(\QGz)$. If the volume
fluctuations are not too large, one can expand the chemical factors $f_{j(n)}$ 
around the mean volume $\overline V$ up to the first order term:

\begin{equation}
f_{j(n)}(V) \simeq f_{j(n)}(\overline V) + f'_{j(n)}(\overline V) (V-\overline V) \; ,
\end{equation}
so that:

\begin{equation}  
 \langle\!\langle n_j \rangle\!\rangle = 
 \sum_{n=1}^{\infty} (\mp 1)^{n+1} \, \gamma_s^{n s_j} \overline V \,
 \xi_{j(n)} f_{j(n)}(\overline V) + 
  \gamma_s^{n s_j} \xi_{j(n)} f'_{j(n)}(\overline V) \int \dint V 
  \rho (V) V (V-\overline V) \; .
\end{equation}    
The first term in the series above gives rise to the formula (38), whilst the second
term can be written as:

\begin{equation}
 \gamma_s^{n s_j} \frac{D^2}{\overline V^2} \overline V \xi_{j(n)} \overline V 
   f'_{j(n)}(\overline V) = \frac{D^2}{\overline V^2} \gamma_s^{n s_j} 
  z_{j(n)}(\overline V) \overline V f'_{j(n)}(\overline V) \; ,
\end{equation}
where $D$ is the dispersion of the distribution $\rho(V)$. This term
is then about a factor $D^2/\overline V^2$ smaller than the leading
term.

\newpage 



 \newpage
 
\section*{Figure captions}
      
\begin{itemize}

\medskip 

\item[\rm Figure 1]
 Behaviour of the global partition function $Z$ as a 
 function of electric charge, baryon number and strangeness, keeping 
 all remaining quantum numbers set to zero, for $T = 170$ MeV, $V = 
 20$ fm$^3$ and $\gamma_s = 0.5$
                 
\item[\rm Figure 2]
 Behaviour of the non-strange baryon chemical factor 
 $Z(0,1,0,0,0)/Z(0,0,0,0,0)$ as a function of volume for different 
 values of the temperature and a fixed value $\gamma_s =0.5$ for the 
 strangeness suppression parameter

\item[\rm Figure 3]  
 Results of hadron multiplicity fits for pp collisions at 
 $\sqrt s =$ 19.4, 23.8  and 26 GeV. Experimental average multiplicities are 
 plotted versus calculated ones. The dashed lines are the quadrant bisectors: 
 well fitted points tend to lie on these lines
 
\item[\rm Figure 4]  
 Residual distributions of hadron multiplicity fits for pp
 collisions at $\sqrt s =$ 19.4, 23.8 and 26 GeV

\item[\rm Figure 5]  
 Results of hadron multiplicity fit for pp collisions 
 at $\sqrt s = 27.4$ GeV. Top: the experimental average multiplicities 
 are plotted versus the calculated ones. The dashed line is the 
 quadrant bisector; well fitted points tend to lie on this line. 
 Bottom: residual distributions

\item[\rm Figure 6]  
 Results of hadron multiplicity fit for \ppb collisions at 
 $\sqrt s =$ 200, 546  and 900 GeV. Experimental average multiplicities are 
 plotted versus calculated ones. The dashed lines are the quadrant bisectors: 
 well fitted points tend to lie on these lines

\item[\rm Figure 7]  
 Residual distributions of hadron multiplicity fits for \ppb
 collisions at $\sqrt s =$ 200, 546 and 900 GeV

\item[\rm Figure 8]  
 Freeze-out temperature values found by fitting hadron 
 abundances in pp, \ppb and \ee collisions \cite{beca2} as a function 
 of centre of mass energy; they are consistent with a constant value 
 over an energy range of about two orders of magnitude. The error bars 
 within horizontal ticks at $\sqrt s = 27.4$ GeV pp collisions and at 
 $\sqrt s = 91.2$ GeV \ee collisions are the fit errors; the overall 
 error bars are the sum in quadrature of the fit error and the systematic
 error related to data set variation (see text)

\item[\rm Figure 9]  
 Strangeness suppression parameters $\gamma_s$ 
 found by fitting hadron abundances in pp, \ppb and \ee 
 collisions \cite{beca2} as a function of centre of mass energy. A slow 
 rise of $\gamma_s$ from 19 to 900 GeV in hadronic collisions may be 
 inferred. At equal centre of mass energy, $\gamma_s$ appears to be 
 definitely lower in pp and \ppb collisions than in \ee collisions.
 The error bars within horizontal ticks at $\sqrt s = 27.4$ GeV pp 
 collisions and at $\sqrt s = 91.2$ GeV \ee collisions are the fit 
 errors; the overall error bars are the sum in quadrature of the fit 
 error and the systematic error related to data set variation (see text) 

\item[\rm Figure 10]  
 Local density and pressure of the hadron gas at 
 freeze-out derived from the fitted parameters as a function of centre of 
 mass energy

\item[\rm Figure 11]  
 Dependence of fitted parameters and primary average 
 multiplicities (top) on the mass cut-off in the hadron mass spectrum
 for \ppb collisions at $\sqrt s =$ 900 GeV

\end{itemize}

\newpage

\begin{figure}[htbp]
\mbox{\epsfig{file=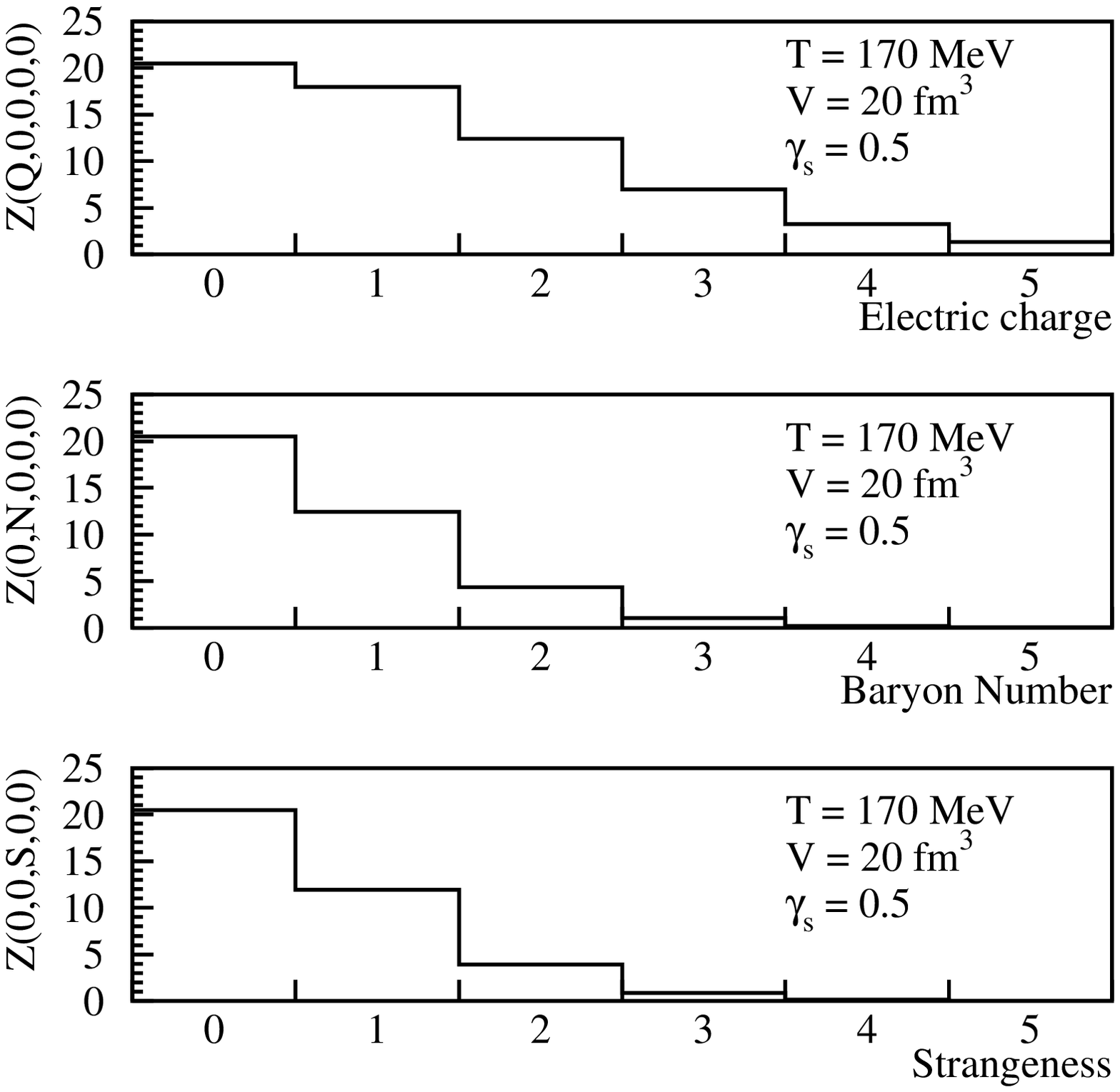,width=17cm}}
\caption{} 
\end{figure}                

\newpage    

\begin{figure}[htbp]
\mbox{\epsfig{file=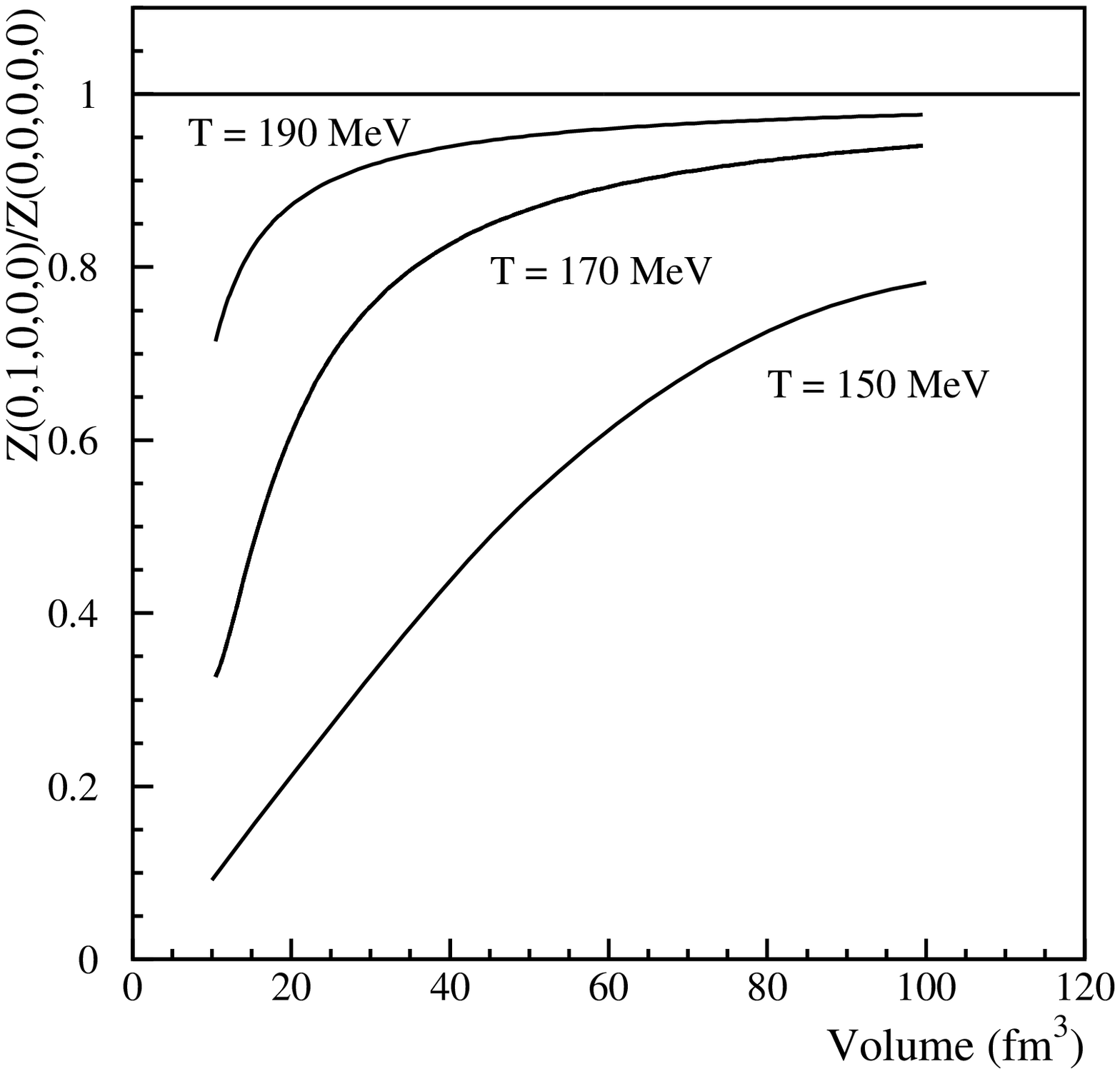,width=17cm}}
\caption{}
\end{figure}  

\newpage  
              
\begin{figure}[htbp]
\mbox{\epsfig{file=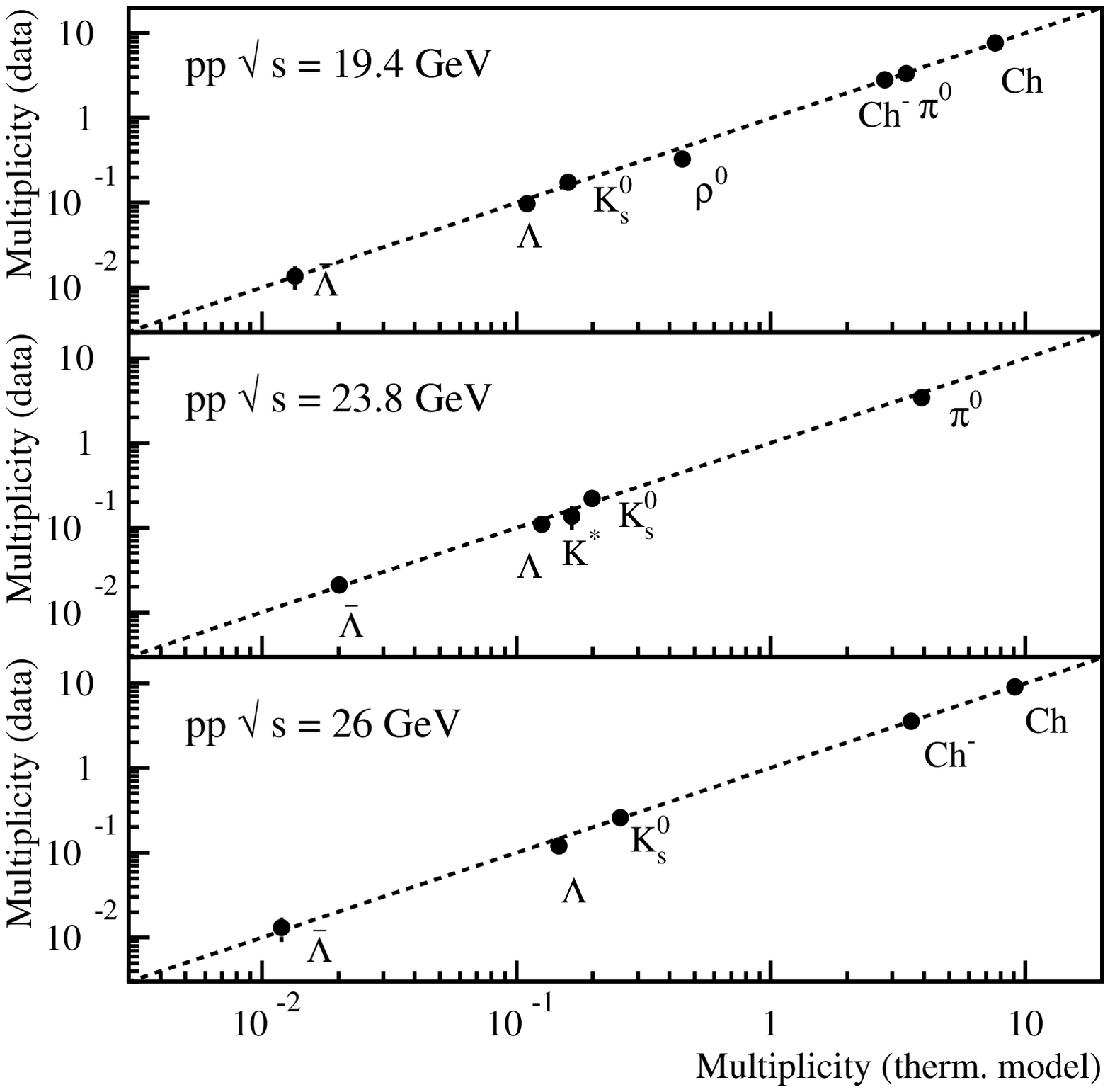,width=17cm}} 
\caption{}
\end{figure}

\newpage  
              
\begin{figure}[htbp]
\mbox{\epsfig{file=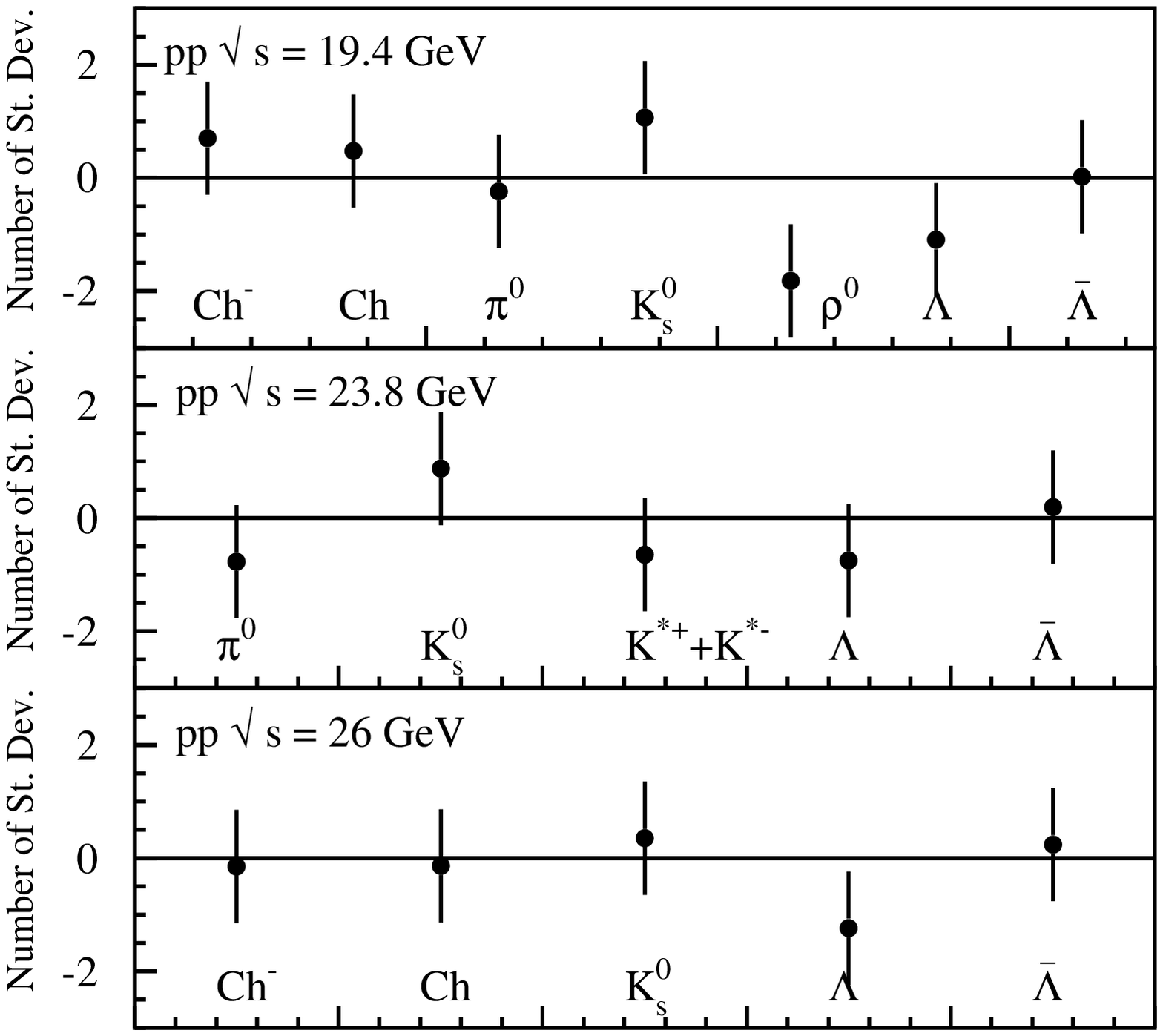,width=17cm}} 
\caption{}
\end{figure}

\newpage  
              
\begin{figure}[htbp]
\mbox{\epsfig{file=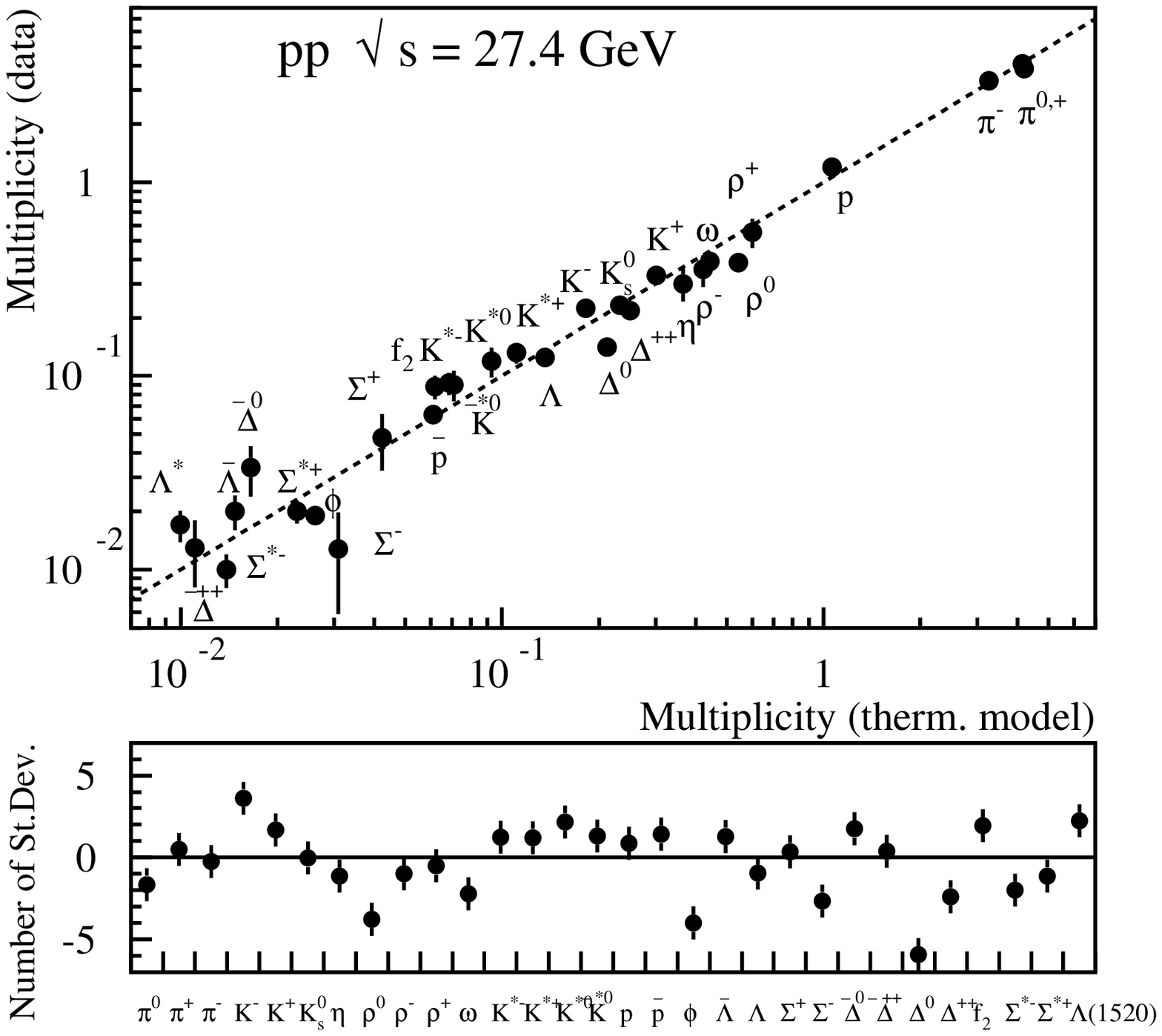,width=17cm}} 
\caption{}
\end{figure}

\newpage  
              
\begin{figure}[htbp]
\mbox{\epsfig{file=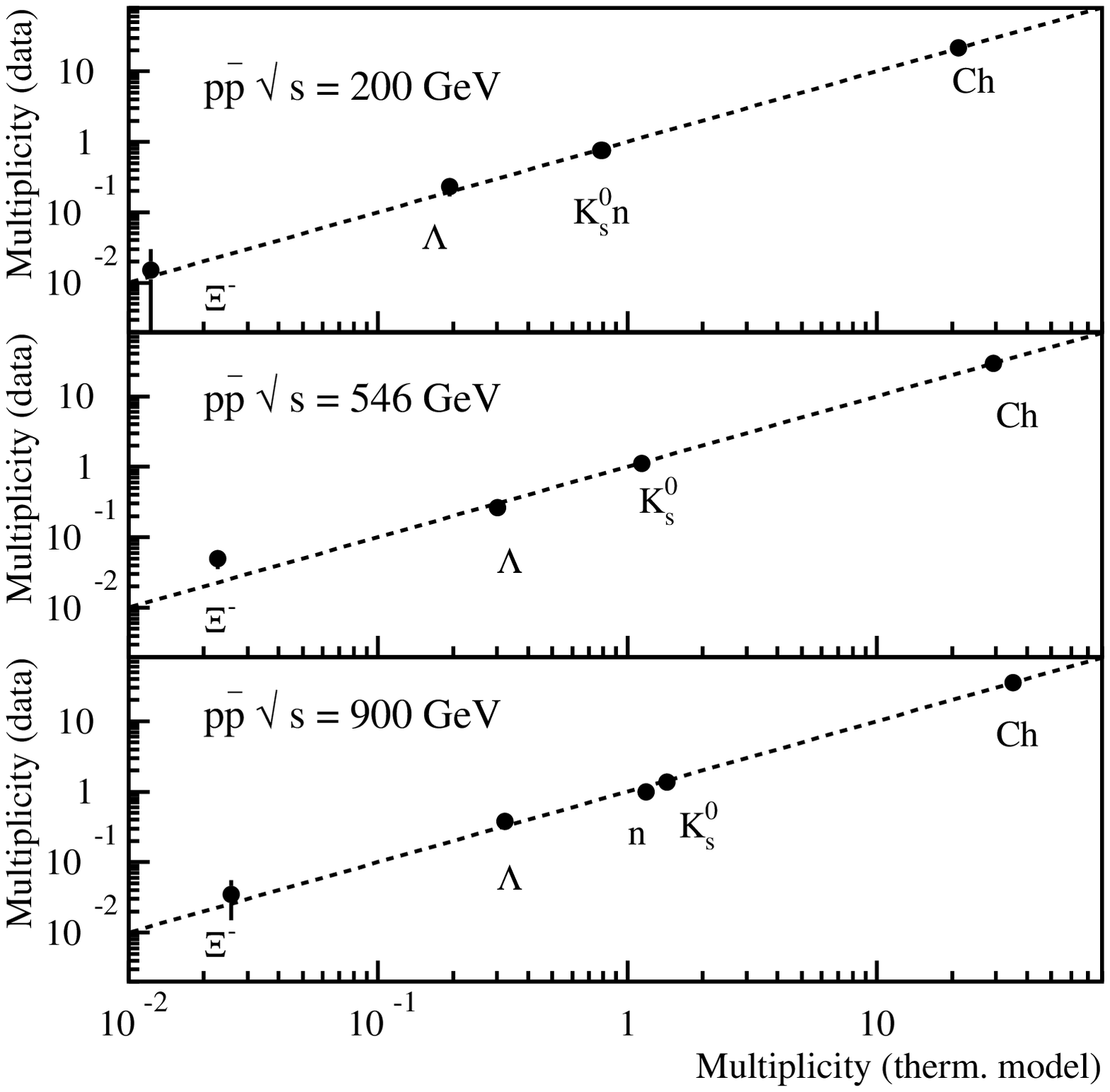,width=17cm}} 
\caption{}
\end{figure}

\newpage  
              
\begin{figure}[htbp]
\mbox{\epsfig{file=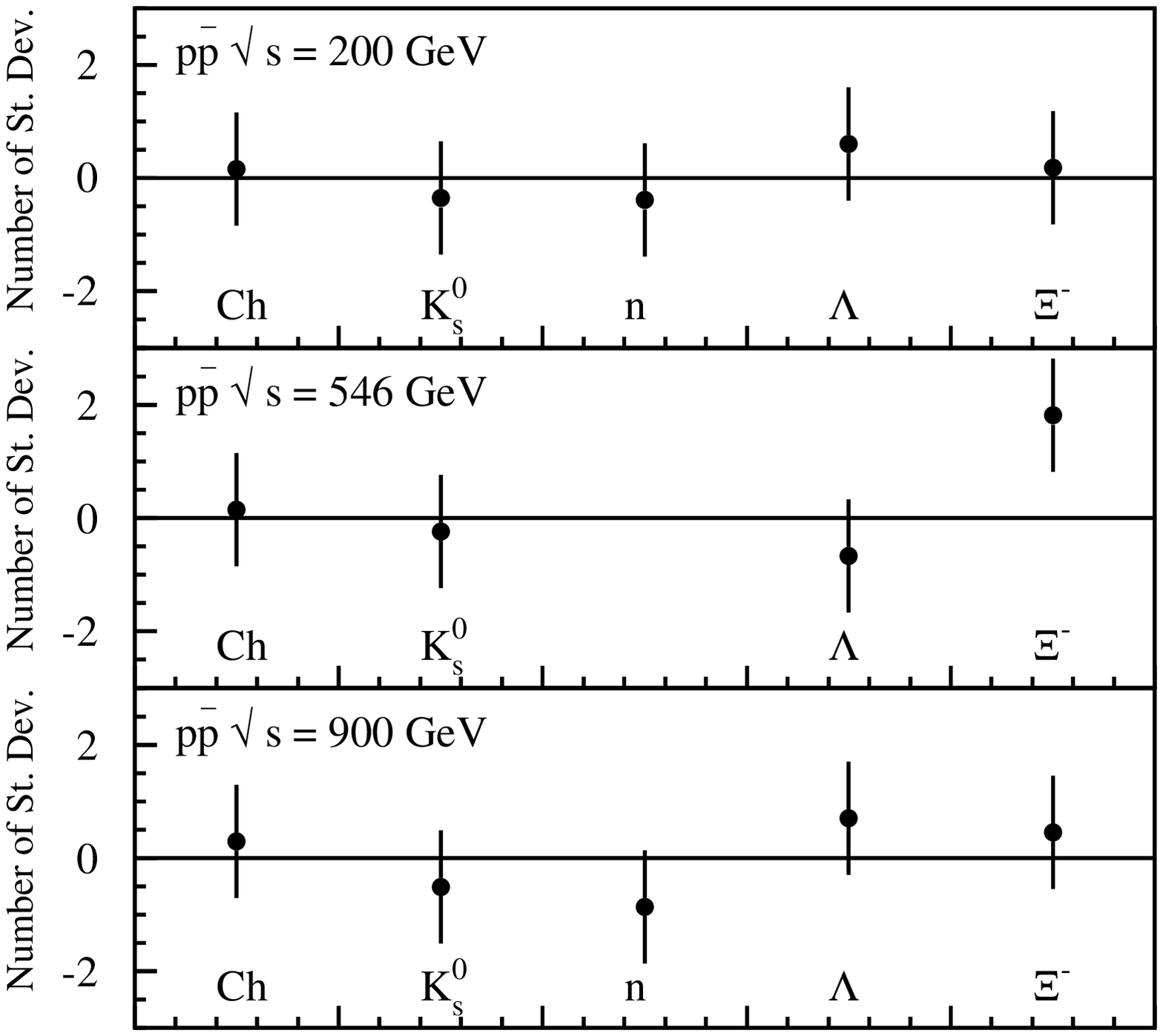,width=17cm}} 
\caption{}
\end{figure}

\newpage  
              
\begin{figure}[htbp]
\mbox{\epsfig{file=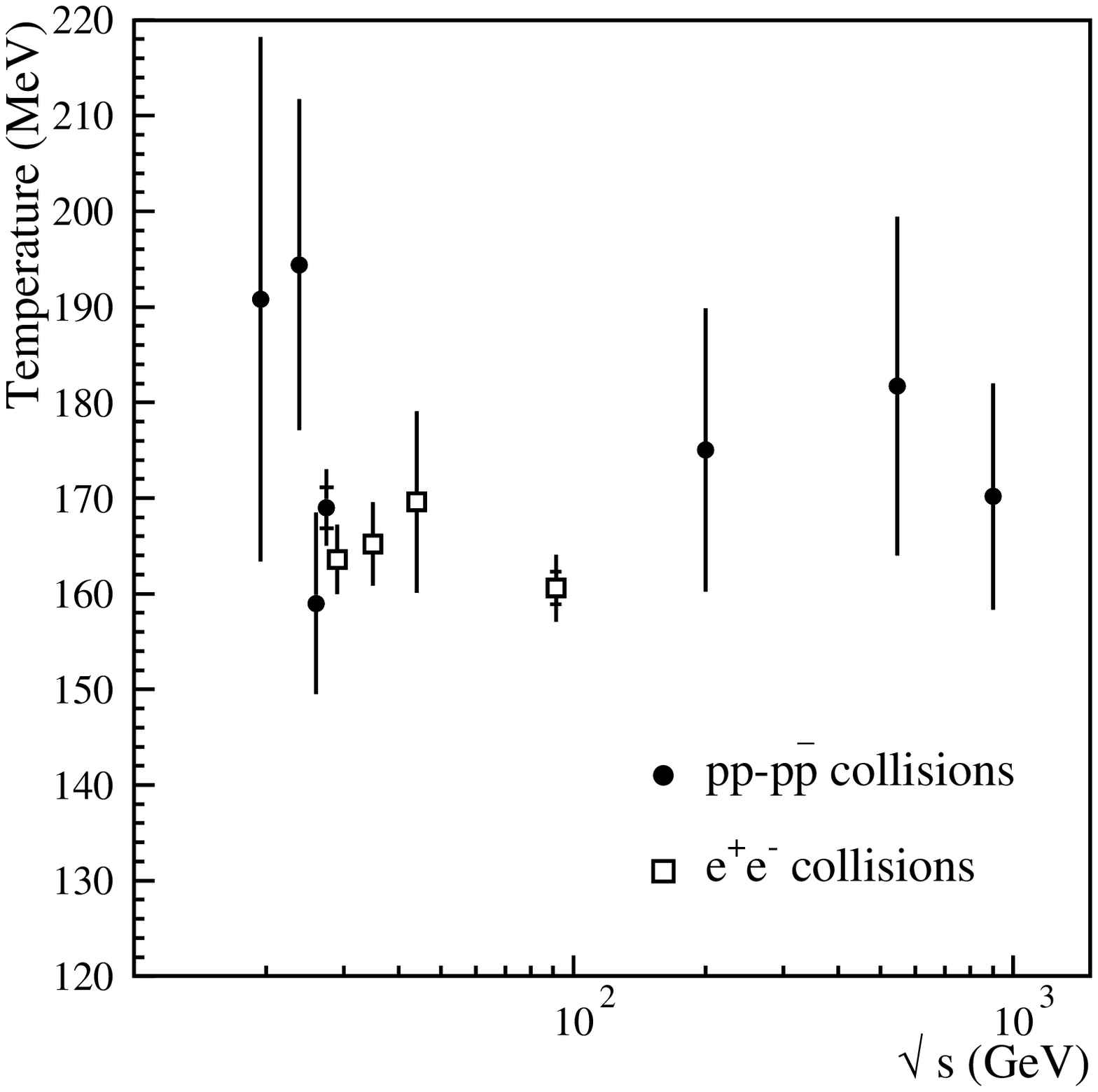,width=17cm}} 
\caption{}
\end{figure}

\newpage  
              
\begin{figure}[htbp]
\mbox{\epsfig{file=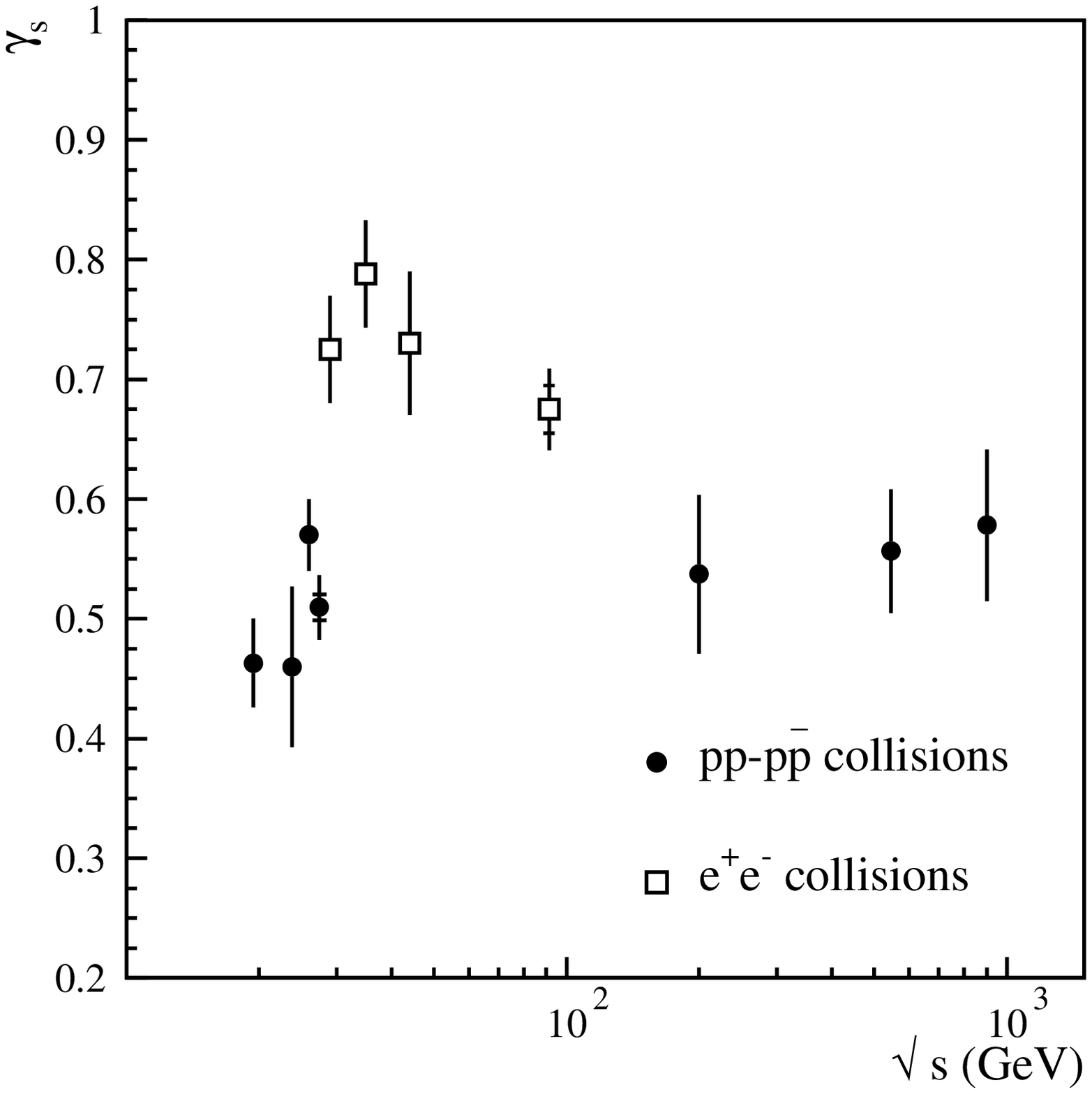,width=17cm}} 
\caption{}
\end{figure}

\newpage  
              
\begin{figure}[htbp]
\mbox{\epsfig{file=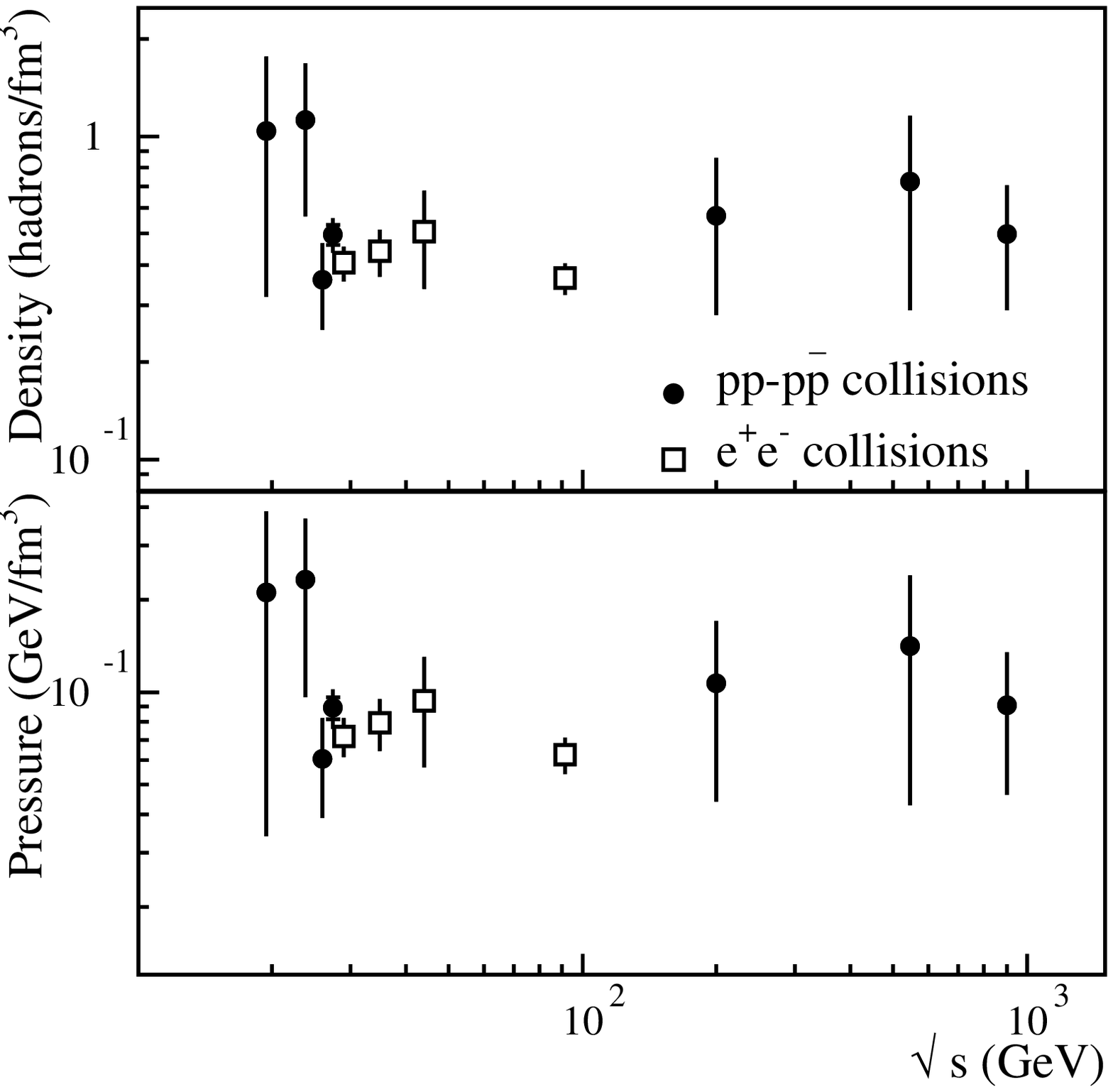,width=17cm}} 
\caption{}
\end{figure}

\newpage  
              
\begin{figure}[htbp]
\mbox{\epsfig{file=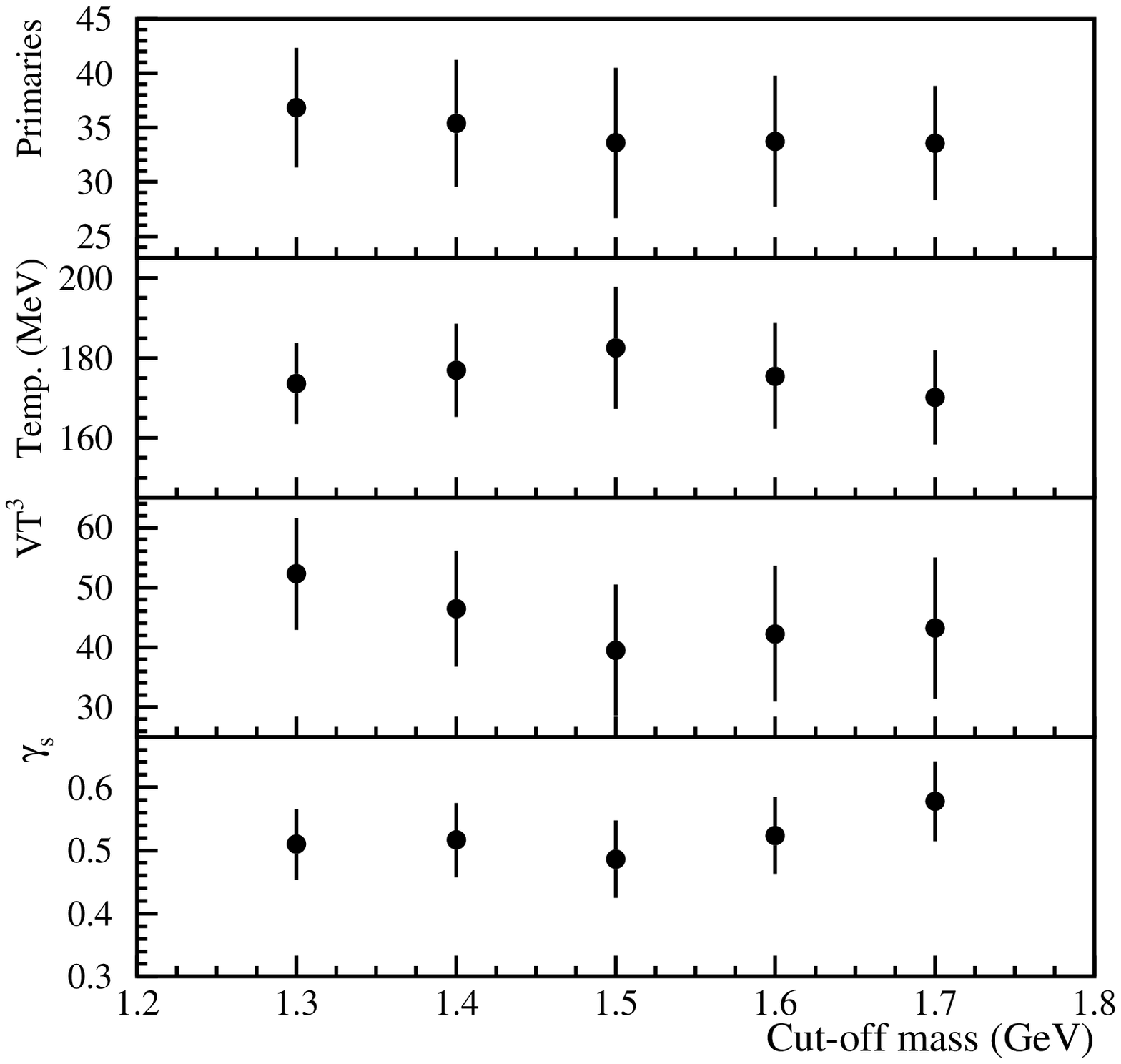,width=17cm}} 
\caption{}
\end{figure}

\end{document}